\documentclass[aps,twocolumn,showpacs,preprintnumbers,floats]{revtex4}\def\@cite#1#2{\textsuperscript{[{#1\if@tempswa , #2\fi}]}}

\usepackage{graphicx}
\usepackage{amsmath}
\usepackage{amsfonts}
\usepackage{amssymb}
\usepackage{color}
\usepackage{subfigure}
\usepackage{epsfig}
\usepackage{morefloats}
\usepackage{multirow}
\usepackage{graphicx,booktabs}
\usepackage{mathrsfs}
\usepackage{txfonts}
\usepackage{indentfirst}
\usepackage{graphicx,booktabs}

\usepackage{longtable,lscape}

\newcommand{\vsig}{\mbox{\boldmath$\sigma$\unboldmath}}
\newcommand{\veps}{\mbox{\boldmath$\epsilon$\unboldmath}}
\newcommand{\vrho}{\mbox{\boldmath$\rho$\unboldmath}}
\newcommand{\vlab}{\mbox{\boldmath$\lambda$\unboldmath}}

\begin{document}

\title{Strong and radiative decays of the low-lying $S$- and $P$-wave singly heavy baryons}
\author{
Kai-Lei Wang$^{1}$, Ya-Xiong Yao$^{1}$, Xian-Hui Zhong$^{1,3}$~\footnote {E-mail: zhongxh@hunnu.edu.cn}, Qiang Zhao$^{2,3,4}$~\footnote {E-mail: zhaoq@ihep.ac.cn}}

\affiliation{ 1) Department
of Physics, Hunan Normal University, and Key Laboratory of
Low-Dimensional Quantum Structures and Quantum Control of Ministry
of Education, Changsha 410081, China }

\affiliation{ 2) Institute of High Energy Physics and Theoretical Physics Center for Science Facilities,
Chinese Academy of Sciences, Beijing 100049, China}

\affiliation{ 3) Synergetic Innovation Center for Quantum Effects and Applications (SICQEA),
Hunan Normal University, Changsha 410081, China}
\affiliation{ 4)  School of Physical Sciences, University of Chinese Academy of Sciences, Beijing 100049, China}

\begin{abstract}

The strong and radiative decays of the low-lying $S$- and $P$-wave $\Lambda_{c(b)}$,
$\Sigma_{c(b)}$, $\Xi_{c(b)}$, $\Xi_{c(b)}'$ and $\Omega_{c(b)}$ baryons
are systematically studied in a constituent quark model. We find that the radiative decay mode $\Lambda_b^0\gamma$ could be very useful for us to establish the missing neutral states $\Sigma_b^{0}$ and $\Sigma_b^{*0}$. Our calculation shows that most of those missing $\lambda$-mode $P$-wave singly heavy baryons have a relatively narrow decay width of less than 30 MeV. Their dominant strong and radiative decay channels can be ideal for searching for their signals in future experiments. The $\Sigma_c(2800)$ resonance may be assigned to $|\Sigma_c~ ^2P_{\lambda} \frac{3}{2}^- \rangle$ with $J^P=3/2^-$ or $|\Sigma_c~ ^4P_{\lambda} \frac{5}{2}^- \rangle$ with $J^P=5/2^-$. In general, the
excitations of $|^2P_{\lambda} \frac{3}{2}^- \rangle$ and $|^4P_{\lambda} \frac{5}{2}^- \rangle$
of the $\mathbf{6}_F$ multiplet have similar strong decay properties. In order to identify them, angular distributions of their decays in either strong decay modes or radiative transitions should be needed.

\end{abstract}

\maketitle

\section{Introduction}{\label{introduction}}

During the past several years, great progress on the heavy baryon spectra has been made in experiments ~\cite{Olive:2016xmw,Chen:2016spr,Cheng:2015iom,Crede:2013sze,Klempt:2009pi} and have provoked many interests in the study of heavy baryon spectroscopy. In the singly charmed baryons, besides the ground states with $J^P=1/2^+$ and $J^P=3/2^+$ ($1S$ wave),
several $P$-wave states, $\Lambda_c(2593)$ ($J^P=1/2^-$),
$\Lambda_c(2625)$ ($J^P=3/2^-$), $\Xi_c(2790)$ ($J^P=1/2^-$), and
$\Xi_c(2815)$ ($J^P=3/2^-$) have been established.
Signals for higher charmed baryons,
such as $\Lambda_c(2880)$~\cite{Artuso:2000xy}, $\Sigma_c(2800)$~\cite{Mizuk:2004yu},
$\Xi_c(2930)$~\cite{Aubert:2007eb}, $\Xi_c(2970,3080)$~\cite{Chistov:2006zj},
$\Xi_c(3055,3123)$~\cite{Aubert:2007dt} and
$\Lambda_c(2860)$~\cite{Aaij:2017vbw}, were also reported by experimental observations.
Very recently, five extremely narrow $\Omega_c(X)$
states, $\Omega_c(3000)$, $\Omega_c(3050)$, $\Omega_c(3066)$, $\Omega_c(3090)$ and $\Omega_c(3119)$,
were observed in the $\Xi_c^{+}K^-$ channel by LHCb~\cite{Aaij:2017nav}.
A detailed summary of the observed charmed baryons can be found in the
recent work by Cheng and Chiang~\cite{Cheng:2017ove}.
On the other hand, in the singly bottom baryons, thanks to the continuous efforts of the LHCb, CDF and CMS~\cite{Basile:1981wr,Aaij:2012da,Aaltonen:2007ar,Aaltonen:2013tta,Abreu:1995kt,
Aaij:2014yka,Chatrchyan:2012ni,Abazov:2008qm,Aaij:2014lxa,Aaij:2016jnn},
most of the ground states with $J^P=1/2^+$ and $J^P=3/2^+$ have been established
except that $\Omega_b^*$ and $\Sigma_b^{0}$ and $\Sigma_b^{*0}$ remain to be found (see Table~\ref{expermentb}).
Recently, two $P$-wave states $\Lambda_b(5912)$ ($J^P=1/2^-$) and $\Lambda_b(5920)$ ($J^P=3/2^-$)
were established by the LHCb experiment~\cite{Aaij:2012da}. While the LHC experiments have demonstrated their discovery capability of heavy flavored baryons, the forthcoming Belle II experiment will also offer another opportunity for the further study of excited heavy baryons.

\begin{table*}[htb]
\begin{center}
\caption{ \label{expermentb} Summary of the observed bottom baryons. Experimental values are taken from the Particle
Data Group (PDG)~\cite{Olive:2016xmw}.}
%\footnotesize
\begin{tabular}{p{1.8cm}p{1.0cm}p{3.2cm}p{3.2cm}p{2.5cm}p{3.0cm}p{2.0cm}ccccccccccc}
\hline\hline
State                     &$J^P$               &Mass (MeV)                           &Width (MeV)                          &Decay channel                 &Experiment             &     Status\\
\hline
$\Lambda_b^0$             &$\frac{1}{2}^+$     &$5619.51\pm0.23$                &$(1466\pm10)\times10^{-15}s$    &$pK^-\pi^+\pi^-$           &CERN R415~\cite{Basile:1981wr}        &***\\
$\Lambda_b(5912)^0$       &$\frac{1}{2}^-$     &$5912.11\pm0.13\pm0.23$         &$<0.66$                         &$\Lambda_b^0\pi^+\pi^-$    &LHCb~\cite{Aaij:2012da}             &***\\
$\Lambda_b(5920)^0$       &$\frac{3}{2}^-$     &$5919.81\pm0.23$                &$<0.63$                         &$\Lambda_b^0\pi^+\pi^-$    &LHCb~\cite{Aaij:2012da},CDF~\cite{Aaltonen:2013tta}             &***\\
$\Sigma_b^{+}$            &$\frac{1}{2}^+$     &$5811.3^{+0.9}_{-0.8}\pm1.7$    &$9.7^{+3.8+1.2}_{-2.8-1.1}$     &$\Lambda_b^0\pi$          &CDF~\cite{Aaltonen:2007ar}           &***  \\
$\Sigma_b^{-}$            &$\frac{1}{2}^+$     &$5815.5^{+0.6}_{-0.5}\pm1.7$    &$4.9^{+3.1}_{-2.1}\pm1.1$       &$\Lambda_b^0\pi$          &CDF~\cite{Aaltonen:2007ar}           &***\\
$\Sigma_b^{*+}$           &$\frac{3}{2}^+$     &$5832.1\pm0.7^{+1.7}_{-1.8}$   &$11.5^{+2.7+1.0}_{-2.2-1.5}$     &$\Lambda_b^0\pi$          &CDF~\cite{Aaltonen:2007ar}           &***\\
$\Sigma_b^{*-}$           &$\frac{3}{2}^+$     &$5835.1\pm0.6^{+1.7}_{-1.8}$   &$7.5^{+2.2+0.9}_{-1.8-1.4}$     &$\Lambda_b^0\pi$          &CDF~\cite{Aaltonen:2007ar}            &***\\
$\Xi_b^{0}$               &$\frac{1}{2}^+$     &$5791.9\pm0.5$              &$(1464\pm31)\times10^{-15}s$      &  $\Xi_c^+\pi^-$                         &DELPHI~\cite{Abreu:1995kt}            &***\\
$\Xi_b^{-}$               &$\frac{1}{2}^+$     &$5794.5\pm1.4$              &$(1560\pm40)\times10^{-15}s$      &  $\Xi_c^0\pi^-$, $J/\psi\Xi^-$              &DELPHI~\cite{Abreu:1995kt}            &***\\
$\Xi_b^{\prime}(5935)^-$  &$\frac{1}{2}^+$     &$5935.02\pm0.02\pm0.05$     &$<0.08$                     &$\Xi_b^0\pi^-$                   &LHCb~\cite{Aaij:2014yka}              &*** \\
$\Xi_b^*(5945)^{0}$       &$\frac{3}{2}^+$     &$5948.9\pm0.8\pm1.4$        &$2.1\pm1.7$                 &$\Xi_b^-\pi^+$                    &CMS~\cite{Chatrchyan:2012ni}         &***\\
$\Xi_b^*(5955)^{-}$       &$\frac{3}{2}^+$     &$5955.33\pm0.12\pm0.05$     &$1.65\pm0.31\pm0.10$                 &$\Xi_b^0\pi^-$          &LHCb~\cite{Aaij:2014yka}              &***\\
$\Omega_b^{-}$            &$\frac{1}{2}^+$     &$6046.4\pm1.9$              &$1570^{+230}_{-200}\times10^{-15}s$ &$J/\psi\Omega^-$         &D{\O}~\cite{Abazov:2008qm}            &***  \\
\hline\hline
\end{tabular}
\end{center}
\end{table*}

Based on the observation of most of the ground state heavy baryons, the interesting topic is to look for the low-lying $P$-wave heavy baryons predicted by the quark model. It should be mentioned that all the established $P$-wave heavy baryons belong to $\bar{\mathbf{3}}_F$, and until now,
no $P$-wave charmed and bottomed states of $\mathbf{6}_F$ have been established (see Table~\ref{sp1}). The newly observed $\Omega_c$
states by the LHCb Collaboration and the previously observed states $\Sigma_c(2800)$
by the Belle Collaboration~\cite{Mizuk:2004yu} and $\Xi_c(2930)$ by the BaBar Collaboration~\cite{Aubert:2007eb}
may be good candidates for some of these missing $P$-wave states. Theoretical calculations of their mass spectra,
decay properties and analysis of their quantum numbers are needed for further establishing them in experiment. Different
theoretical model calculations of the mass spectra of the singly heavy
baryons can be found in the literature~\cite{Copley:1979wj,Ebert:2011kk,Ebert:2007nw,Ebert:2005xj,Maltman:1980er,
Yoshida:2015tia,Valcarce:2008dr,Shah:2016mig,Roberts:2007ni,
Shah:2016nxi,Karliner:2008sv,Padmanath:2017lng,Padmanath:2013bla,Bali:2015lka,
Chen:2014nyo,Chen:2016iyi,Lu:2016ctt,Chen:2015kpa,Mao:2015gya,Wang:2017vnc, Wang:2017zjw,Karliner:2017kfm,Aliev:2017led,Gerasyuta:2007un,Garcilazo:2007eh}.
Except for the mass calculations, the systematical predictions of the decays of the
$P$-wave heavy baryons are urgently needed in theory. For this
purpose, in the present work, we carry out a systematic study of the strong
and radiative decays of the low-lying $P$-wave heavy baryons. Furthermore,
to provide information for the last three missing $1S$ states $\Omega_b^{*-}$,
$\Sigma_b^{0}$, and $\Sigma_b^{*0}$, and to better understand
the properties of the $1S$ singly heavy baryons,
their strong and radiative decays are studied as well.

In this work we apply the chiral quark model (ChQM)~\cite{Manohar:1983md} to deal with the strong decays of the singly heavy baryons.
In this framework, the spatial wave functions of heavy baryons
are described by harmonic oscillators, and an effective chiral Lagrangian is
then introduced to account for the quark-meson coupling at the baryon-meson interaction vertex.
The light pseudoscalar mesons, i.e., $\pi$, $K$, and $\eta$,
are treated as Goldstone bosons. Since the quark-meson coupling is invariant under the
chiral transformation, some of the low-energy properties
of QCD are retained~\cite{Manohar:1983md,Li:1997gd,Zhao:2002id}. This model has been developed
and successfully used to deal with the strong decays of heavy-light mesons, charmed
and strange baryons~\cite{Zhong:2008kd,Zhong:2010vq,Zhong:2009sk,
Zhong:2007gp,Liu:2012sj,Xiao:2013xi,Nagahiro:2016nsx,Wang:2017hej,Xiao:2014ura,Xiao:2017udy}.
This approach is different from the often used $^3P_0$ model~\cite{Micu:1968mk,LeYaouanc:1972vsx,
LeYaouanc:1973ldf,Chen:2007xf,Chen:2017gnu,Zhao:2017fov,Ye:2017yvl} since different effective degrees of freedom are involved.
In the ChQM the light pseudoscalar mesons behave as pointlike particles which couple to the light constituent quark within the baryons via the effective chiral Lagrangian. In contrast, the transition operator in the $^3P_0$ model is obtained by a quark pair creation with the vacuum quantum numbers $J^P=0^{++}$. The created pair of quarks will rearrange with the quarks within the initial baryon to the final state meson and baryon. Thus, the hadronic transition amplitude will be described by operators extracted at the constituent quark level.
Many other model studies of the strong decays of the low-lying $S$- and $P$-wave
heavy baryons can be found in the literature.
For example, for the singly charmed baryons, the strong decay properties of
the low-lying $S$- and/or $P$-wave states were studied with the method of
light cone QCD sum rules (LCQSR)~\cite{Chen:2017sci,Zhu:2000py,Agaev:2017lip,Agaev:2017nn},
the $^3P_0$ model~\cite{Chen:2007xf,Chen:2017gnu,Zhao:2017fov,Ye:2017yvl}, the heavy hadron chiral perturbation
theory (HHChPT)~\cite{Cheng:2006dk,Cho:1994vg,Pirjol:1997nh,
Chiladze:1997ev,Cheng:2015iom,Blechman:2003mq,Huang:1995ke}, the light front quark model
~\cite{Tawfiq:1998nk,Tawfiq:1999vz}, the relativistic three-quark model (RQM)~\cite{Ivanov:1999bk,Ivanov:1998qe,
Korner:1994nh,Hussain:1999sp},
the nonrelativistic quark model (NQM)~\cite{Albertus:2005zy}, the MIT bag model~\cite{Hwang:2006df},
and the Bethe-Salpeter formalism~\cite{Guo:2007qu}. For the strong decay properties of the singly bottom baryons,
only a few studies are found in the literature~\cite{Hernandez:2011tx,Guo:2007qu,
Hwang:2006df,Liu:2012sj,Chen:2007xf,Limphirat:2010zz}. While most of the discussions
focus on the $S$-wave ground states, a systematic study of the strong decays of the $P$-wave singly bottom
baryons seems to be needed.

For the radiative transitions, we exploit an EM transition operator which is extracted in the nonrelativistic constituent quark model and has been successfully applied to the study of the radiative decays of $c\bar{c}$ and $b\bar{b}$ systems
~\cite{Deng:2016stx,Deng:2016ktl} and the heavy baryons~\cite{Wang:2017hej,Lu:2017meb,Xiao:2017udy}.
There are some discussions of the radiative decays of the singly heavy baryons in the literature~\cite{Cheng:1992xi,Wang:2009ic,Wang:2009cd,Jiang:2015xqa,Zhu:1998ih,Tawfiq:1999cf,Dey:1994qi,
Bernotas:2013eia,Gamermann:2010ga,Aliev:2014bma,Aliev:2009jt,Aliev:2016xvq,Aliev:2011bm,
Chow:1995nw,Bahtiyar:2016dom,Bahtiyar:2015sga,Ivanov:1998wj,Savage:1994wa,Banuls:1999br}. While most of the studies focus on the $S$-wave ground states, only a few studies considered the $P$-wave heavy baryons EM transitions~\cite{Cho:1994vg,Ivanov:1999bk,Ivanov:1998wj,Zhu:2000py,Tawfiq:1999cf,Gamermann:2010ga,Chow:1995nw}.
Theoretical studies of the radiative transitions by the lattice QCD~\cite{Bahtiyar:2016dom},  LCQSR~\cite{Zhu:1998ih,Wang:2009ic,Wang:2009cd,Aliev:2014bma,Aliev:2009jt,Aliev:2016xvq},
HHChPT~\cite{Cheng:1992xi,Cho:1994vg,Jiang:2015xqa,Banuls:1999br,Savage:1994wa},
RQM~\cite{Ivanov:1999bk,Ivanov:1998wj}, the bag model (BM)~\cite{Bernotas:2013eia},
the vector meson dominance model (VMD)~\cite{Aliev:2011bm}, and the model based on heavy quark
symmetry (HQS)~\cite{Tawfiq:1999cf}, can also be found in the literature.
It should be mentioned that there appears a strong model-dependence in various model calculations
which indicates that the radiative transition mechanism for these heavy baryons still need to be understood.

As follows, in Sec.~\ref{spectrum} we first give a brief description of the heavy baryon spectra.
Then a brief introduction to the quark model treatment for the strong and radiative decays of
the singly heavy baryon system will be given in Sec.~\ref{decays}.
The numerical results are presented and discussed in Sec.~\ref{results}. Finally, a summary is
given in Sec.\ \ref{suma}.

\section{heavy baryon spectrum}\label{spectrum}

The heavy baryon containing a heavy quark violates
the SU(4) symmetry. However, the SU(3) symmetry between the other two
light quarks ($u$, $d$, or $s$) is approximately kept. According to
the symmetry, the heavy baryons containing a single heavy quark
belong to two different SU(3) flavor representations: the symmetric sextet $\mathbf{6}_F$ and
antisymmetric antitriplet $\bar{\mathbf{3}}_F$. For the charmed
baryons $\Lambda_c$ and $\Xi_c$ belonging to $\bar{\mathbf{3}}_F$, the antisymmetric flavor
wave functions can be written as
\begin{equation} \phi^c_{\bar{\mathbf{3}}}=\begin{cases}
        \frac{1}{\sqrt{2}}(ud-du)c   &$for$~\Lambda_c^{+},\\
        \frac{1}{\sqrt{2}}(us-su)c   &$for$~\Xi_c^+,\\
        \frac{1}{\sqrt{2}}(ds-sd)c   &$for$~\Xi_c^0;
       \end{cases}
\end{equation}
and for the bottom baryons $\Lambda_b$ and $\Xi_b$ belonging to $\bar{\mathbf{3}}_F$, the antisymmetric flavor
wave functions can be written as
\begin{equation} \phi^b_{\bar{\mathbf{3}}}=\begin{cases}
        \frac{1}{\sqrt{2}}(ud-du)b   &$for$~\Lambda_b^{0},\\
        \frac{1}{\sqrt{2}}(us-su)b   &$for$~\Xi_b^0,\\
        \frac{1}{\sqrt{2}}(ds-sd)b   &$for$~\Xi_b^-.
       \end{cases}
\end{equation}
For the charmed baryons belonging to $\mathbf{6}_F$, the symmetric flavor
wave functions can be written as
\begin{equation}
\phi^c_{\mathbf{6}}=\begin{cases}
                             uuc     &$for$~\Sigma^{++}_c,\\
        \frac{1}{\sqrt{2}}(ud+du)c   &$for$~\Sigma_c^{+},\\
                             ddc     &$for$~\Sigma^{0}_c,\\
        \frac{1}{\sqrt{2}}(us+su)c   &$for$~\Xi_c^{'+},\\
        \frac{1}{\sqrt{2}}(ds+sd)c   &$for$~\Xi_c^{'0},\\
                               ssc   &$for$~\Omega_c^0;
       \end{cases}
\end{equation}
while for the bottom baryons belonging to $\mathbf{6}_F$, the symmetric flavor
wave functions can be written as
\begin{equation}
\phi^b_{\mathbf{6}}=\begin{cases}
                             uub     &$for$~\Sigma^{+}_b,\\
        \frac{1}{\sqrt{2}}(ud+du)b   &$for$~\Sigma_b^{0},\\
                             ddb     &$for$~\Sigma^{-}_b,\\
        \frac{1}{\sqrt{2}}(us+su)b   &$for$~\Xi_b^{'0},\\
        \frac{1}{\sqrt{2}}(ds+sd)b   &$for$~\Xi_b^{'-},\\
                               ssb   &$for$~\Omega_b^-;
       \end{cases}
\end{equation}

In the quark model, the typical SU(2) spin wave functions of the
heavy baryons can be adopted:
\begin{eqnarray}
\chi^s_{3/2}&=&\uparrow\uparrow\uparrow, \ \  \chi^s_{-3/2}=\downarrow\downarrow\downarrow, \nonumber\\
\chi^s_{1/2}&=&\frac{1}{\sqrt{3}}(\uparrow\uparrow\downarrow+\uparrow\downarrow\uparrow+\downarrow\uparrow\uparrow),\nonumber\\
\chi^s_{-1/2}&=&\frac{1}{\sqrt{3}}(\uparrow\downarrow\downarrow+\downarrow\downarrow\uparrow+\downarrow\uparrow\downarrow),
\end{eqnarray}
for the spin-3/2 states with symmetric spin wave functions;
\begin{eqnarray}
\chi^\rho_{1/2}&=&\frac{1}{\sqrt{2}}(\uparrow\downarrow\uparrow-\downarrow\uparrow\uparrow),\nonumber\\
\chi^\rho_{-1/2}&=&\frac{1}{\sqrt{2}}(\uparrow\downarrow\downarrow-\downarrow\uparrow\downarrow),
\end{eqnarray}
for the spin-1/2 states with mixed antisymmetric spin wave
functions; and
\begin{eqnarray}
\chi^\lambda_{1/2}&=&-\frac{1}{\sqrt{6}}(\uparrow\downarrow\uparrow+\downarrow\uparrow\uparrow-2\uparrow\uparrow\downarrow),\nonumber\\
\chi^\lambda_{-1/2}&=&+\frac{1}{\sqrt{6}}(\uparrow\downarrow\downarrow+\downarrow\uparrow\downarrow-2\downarrow\downarrow\uparrow),
\end{eqnarray}
for the spin-1/2 states with mixed symmetric spin wave functions.

The spatial wave function of a heavy baryon is adopted from the
harmonic oscillator form in the constituent quark model~\cite{Zhong:2007gp}.
For a $q_1q_2Q$ basis state, it contains
two light quarks $q_1$ and $q_2$ with
an equal mass $m$, and a heavy quark $Q$ with mass $m'$. The
basis states are generated by the oscillator Hamiltonian
\begin{eqnarray} \label{hm2}
\mathcal{H}=\frac{P^2_{cm}}{2 M}+\frac{1}{2m_\rho}\mathbf{p}^2_\rho+\frac{1}{2m_\lambda}\mathbf{p}^2_\lambda+
\frac{3}{2}K(\rho^2+\lambda^2).
\end{eqnarray}
The constituent quarks are confined in an oscillator potential with the potential parameter $K$ independent
of the flavor quantum number. The Jacobi coordinates $\vrho$ and $\vlab$ and center-of-mass (c.m.) coordinate
$\mathbf{R}_{c.m.}$ can be related to the coordinate $\textbf{r}_{j}$ of the $j$th
quark by
\begin{eqnarray}
\vrho&=&\frac{1}{\sqrt{2}}(\mathbf{r}_1-\mathbf{r}_2),\label{zb1}\\
\vlab&=&\frac{1}{\sqrt{6}}(\mathbf{r}_1+\mathbf{r}_2-2\mathbf{r}_3),\label{zb2}\\
\mathbf{R}_{c.m.}&=&\frac{m(\mathbf{r}_1+\mathbf{r}_2)+m'\mathbf{r}_3}{2m+m'}\label{zb3},
\end{eqnarray}
and the momenta $\mathbf{p}_\rho$, $\mathbf{p}_\lambda $ and $\mathbf{P}_{c.m.}$ are defined by
\begin{eqnarray} \label{mom}
\mathbf{p}_\rho=m_\rho\dot{\vrho},\ \
\mathbf{p}_\lambda=m_\lambda\dot{\vlab},\ \
\mathbf{P}_{c.m.}=M \mathbf{\dot{R}}_{c.m.},
\end{eqnarray}
with
\begin{eqnarray} \label{mass}
M=2m+m',\ \ m_\rho=m,\ \ m_\lambda=\frac{3m m'}{2m+m'}.
\end{eqnarray}
The wave function of an oscillator is given by
\begin{eqnarray}
\psi^{n_\sigma}_{l_\sigma m}(\sigma)=R_{n_\sigma
l_\sigma}(\sigma)Y_{l_\sigma m}(\sigma),
\end{eqnarray}
where $\sigma=\rho,\lambda$.
In the wave functions, there are two oscillator parameters, i.e. the
potential strengths $\alpha_\rho$ and $\alpha_\lambda$.
The parameters $\alpha_\rho$ and $\alpha_\lambda$
satisfy the following relation~\cite{Zhong:2007gp}:
\begin{eqnarray} \label{rhol}
\alpha^2_\lambda=\sqrt{\frac{3m'}{2m+m'}}\alpha^2_\rho.
\end{eqnarray}
The spatial wave function is a product of the $\rho$-oscillator
and the $\lambda$-oscillator states. With the standard notation,
the principal quantum numbers of the $\rho$-mode oscillator and
$\lambda$-mode oscillator are $N_\rho=(2n_\rho+l_\rho)$ and
$N_\lambda=(2n_\lambda+l_\lambda)$, and  the energy of a state is
given by
\begin{eqnarray}
E&=&(N_\rho+\frac{3}{2})\omega_\rho+(N_\lambda+\frac{3}{2})\omega_\lambda \ .
\end{eqnarray}
with the $\rho$-mode and $\lambda$-mode frequencies
\begin{eqnarray}\label{freq}
\omega_\rho=(3K/m_\rho)^{1/2},\ \
\omega_\lambda=(3K/m_\lambda)^{1/2}.
\end{eqnarray}
Combining the relation $\omega_\rho=\sqrt{3m'/(2m+m')} \omega_\lambda >\omega_\lambda$,
we find that if we take $N_\rho=N_\lambda$, the $\rho$-mode excited energy
$(N_\rho+\frac{3}{2})\omega_\rho$ is much larger than
the $\lambda$-mode excited energy $(N_\lambda+\frac{3}{2})\omega_\lambda$,
which indicates that the $\lambda$-mode excitations should be more
easily formed than the $\rho$-mode excitations.
Thus, in present work we only study the $\lambda$-mode excitations.

The product of spin, flavor, and spatial wave functions of the heavy baryons
must be symmetric since the color wave function is antisymmetric. More details
about the classifications of the heavy baryons in the quark model
can be found in our previous work~\cite{Zhong:2007gp}.
The predicted mass spectra of the $1S$-wave and $\lambda$-mode $1P$-wave single
heavy baryons within various quark models are summarized in Tab.~\ref{sp1}.

%p{1.6cm}|p{1.8cm}p{1.8cm}p{1.8cm}p{1.8cm}p{1.8cm}p{1.8cm}p{1.8cm}p{1.8cm}p{1.8cm}
\begin{table*}[htp]
\begin{center}
\caption{\label{sp1}  Mass spectra of the low-lying $S$- and $P$-wave singly heavy baryons from
various quark models~\cite{Ebert:2011kk,Yoshida:2015tia,Chen:2016iyi,Roberts:2007ni} compared
with the data from the PDG~\cite{Olive:2016xmw}.}
\scalebox{1.0}{
\begin{tabular}{c|ccccccccccccccccccccccccccccccccccccccccccccc}\hline\hline
&\multicolumn{4}{c}{$\underline{~~~~~~~~~~~~~~~~~~~~~~~~~~~~~~~~~~~~~~~~~~~~~~~~~~~~~~~~~\Lambda_c~~~~~~~~~~~~~~~~~~~~~~~~~~~~~~~~~~~~~~~}$}    &\multicolumn{3}{c}{$\underline{~~~~~~~~~~~~~~~~~~~~~~~~~~~~~~~~~~\Lambda_b ~~~~~~~~~~~~~~~~~~~~~~~~~~~}$}\\
State          &~~~~RQM~\cite{Ebert:2011kk}~~~~&~~~~NQM~\cite{Yoshida:2015tia}~~~~    &~~~~NQM~\cite{Chen:2016iyi}~~~~      &~~~~PDG~\cite{Olive:2016xmw}~~~~   &~~~~RQM~\cite{Ebert:2011kk}~~~~ &~~~~NQM~\cite{Yoshida:2015tia}~~~~ &~~~~PDG~\cite{Olive:2016xmw}~~~~\\ \hline
$1^2S \frac{1}{2}^+$                &2286     &2285        &2286      &2286    &5620      &5618      &5620 \\
$1^2P_{\lambda} \frac{1}{2}^-$       &2598     &2628       &2614      &2592     &5930      &5938      &5912\\
$1^2P_{\lambda} \frac{3}{2}^-$       &2627     &2630       &2639      &2628     &5942      &5939      &5920\\ \hline
&\multicolumn{4}{c}{$\underline{~~~~~~~~~~~~~~~~~~~~~~~~~~~~~~~~~~~~~~~~~~~~~~~~~~~~~~~~~\Xi_c~~~~~~~~~~~~~~~~~~~~~~~~~~~~~~~~~~~~~~~}$}    &\multicolumn{3}{c}{$\underline{~~~~~~~~~~~~~~~~~~~~~~~~~~~~~~~~~~\Xi_b ~~~~~~~~~~~~~~~~~~~~~~~~~~~}$}\\
State          &RQM~\cite{Ebert:2011kk}        &NQM~\cite{Roberts:2007ni}      &NQM~\cite{Chen:2016iyi}     &PDG~\cite{Olive:2016xmw}  &RQM~\cite{Ebert:2011kk}          &NQM~\cite{Roberts:2007ni} &PDG~\cite{Olive:2016xmw}\\ \hline
$1^2S \frac{1}{2}^+$                 &2476     &2466       &2470     &$2471$               &5803     &5806         &5795\\
$1^2P_{\lambda} \frac{1}{2}^-$       &2792     &2773       &2793     &2790                 &6120     &6090         &?\\
$1^2P_{\lambda} \frac{3}{2}^-$       &2819     &2783       &2820     &2815                 &6130     &6093         &?\\
\hline
&\multicolumn{4}{c}{$\underline{~~~~~~~~~~~~~~~~~~~~~~~~~~~~~~~~~~~~~~~~~~~~~~~~~~~~~~~~\Sigma_c~~~~~~~~~~~~~~~~~~~~~~~~~~~~~~~~~~~~~~~~}$}    &\multicolumn{3}{c}{$\underline{~~~~~~~~~~~~~~~~~~~~~~~~~~~~~~~~~~\Sigma_b ~~~~~~~~~~~~~~~~~~~~~~~~~~~}$}\\
State          &RQM~\cite{Ebert:2011kk}       &NQM~\cite{Yoshida:2015tia}   &NQM~\cite{Chen:2016iyi}  &PDG~\cite{Olive:2016xmw} &RQM~\cite{Ebert:2011kk}       &NQM~\cite{Yoshida:2015tia} &PDG~\cite{Olive:2016xmw}\\ \hline
$1^2S \frac{1}{2}^+$                &2443      &2460       &2456     &2455       &5808      &5823          &5811 \\
$1^4S \frac{3}{2}^+$                &2519      &2523       &2515     &2520       &5834       &5845         &5832 \\
$1^2P_{\lambda} \frac{1}{2}^-$      &2713      &2802       &2702     & ?          &6101       &6127        &?\\
$1^2P_{\lambda} \frac{3}{2}^-$      &2798      &2807       &2785     & ?          &6096       &6132 &?\\
$1^4P_{\lambda} \frac{1}{2}^-$      &2799      &2826       &2765     & ?          &6095       &6135 &?\\
$1^4P_{\lambda} \frac{3}{2}^-$      &2773      &2837       &2798     & ?          &6087       &6141 &?\\
$1^4P_{\lambda} \frac{5}{2}^-$      &2789      &2839       &2790     & ?          &6084       &6144 &?\\
\hline
&\multicolumn{4}{c}{$\underline{~~~~~~~~~~~~~~~~~~~~~~~~~~~~~~~~~~~~~~~~~~~~~~~~~~~~~~~~\Xi^{\prime}_c~~~~~~~~~~~~~~~~~~~~~~~~~~~~~~~~~~~~~~~~}$}    &\multicolumn{3}{c}{$\underline{~~~~~~~~~~~~~~~~~~~~~~~~~~~~~~~~~~\Xi^{\prime}_b ~~~~~~~~~~~~~~~~~~~~~~~~~~~}$}\\
State          &RQM~\cite{Ebert:2011kk}       & NQM~\cite{Roberts:2007ni}   &NQM~\cite{Chen:2016iyi}     &PDG~\cite{Olive:2016xmw}   &RQM~\cite{Ebert:2011kk}  & NQM~\cite{Roberts:2007ni}&PDG~\cite{Olive:2016xmw}               \\ \hline
$1^2S \frac{1}{2}^+$                 &2579     & 2592              &2579     &2578  &5936  &5958 &5935                  \\
$1^4S \frac{3}{2}^+$                 &2649     & 2650              &2649     &2645  &5963  &5982 &5945(5955)            \\
$1^2P_{\lambda} \frac{1}{2}^-$       &2936     & 2859              &2839     & ?     &6233  &6192 &  ?                     \\
$1^2P_{\lambda} \frac{3}{2}^-$       &2935     & 2871              &2921     & ?     &6234  &6194 & ?                     \\
$1^4P_{\lambda} \frac{1}{2}^-$       &2854     & $\cdot\cdot\cdot$ &2900     & ?     &6227  &$\cdot\cdot\cdot$& ?                     \\
$1^4P_{\lambda} \frac{3}{2}^-$       &2912     &$\cdot\cdot\cdot$  &2932     & ?     &6224  &$\cdot\cdot\cdot$& ?                 \\
$1^4P_{\lambda} \frac{5}{2}^-$       &2929     &2905               &2927     & ?     &6226  &6204 &   ?        \\
\hline
&\multicolumn{4}{c}{$\underline{~~~~~~~~~~~~~~~~~~~~~~~~~~~~~~~~~~~~~~~~~~~~~~~~~~~~~~~~\Omega_c~~~~~~~~~~~~~~~~~~~~~~~~~~~~~~~~~~~~~~~~}$}    &\multicolumn{3}{c}{$\underline{~~~~~~~~~~~~~~~~~~~~~~~~~~~~~~~~~~\Omega_b ~~~~~~~~~~~~~~~~~~~~~~~~~~~}$}\\
State          &RQM~\cite{Ebert:2011kk}  &NQM~\cite{Yoshida:2015tia} &NQM~\cite{Roberts:2007ni}     &PDG~\cite{Olive:2016xmw}    &RQM~\cite{Ebert:2011kk}     &NQM~\cite{Yoshida:2015tia} &PDG~\cite{Olive:2016xmw}\\ \hline
$1^2S \frac{1}{2}^+$                 &2698     &2731   &2718             &2695       &6064       &6076        &6046            \\
$1^4S \frac{3}{2}^+$                 &2768     &2779   &2776             &2770       &6088       &6094       &   ?             \\
$1^2P_{\lambda} \frac{1}{2}^-$       &2966     &3030   &2977             &  ?        &6330       &6333      &   ? \\
$1^2P_{\lambda} \frac{3}{2}^-$       &3029     &3033   &2986             &  ?        &6331       &6336     &   ? \\
$1^4P_{\lambda} \frac{1}{2}^-$       &3055     &3048   &2990             &   ?       &6339       &6340   &   ?  \\
$1^4P_{\lambda} \frac{3}{2}^-$       &3054     &3056   &2994             &   ?       &6340       &6344   &   ? \\
$1^4P_{\lambda} \frac{5}{2}^-$       &3051     &3057   &3014             &   ?       &6334       &6345   &   ? \\
\hline\hline
\end{tabular}}
\end{center}
\end{table*}

\section{Models for strong and radiative decays }\label{decays}

%We apply the chiral quark model~\cite{Manohar:1983md}
%to the study of the hadronic decays of the low-lying charmed and bottom
%baryons. By treating the light pseudoscalar mesons, i.e. $\pi$, $K$ and $\eta$, as Goldstone boson, this method has been successfully applied to the %hadronic decays of  heavy-light mesons, charmed
%and strange baryons~\cite{Zhong:2008kd,Zhong:2010vq,Zhong:2009sk,
%Zhong:2007gp,Liu:2012sj,Xiao:2013xi,Nagahiro:2016nsx,Wang:2017hej,Xiao:2014ura,Xiao:2017udy}.
In the chiral quark model the effective quark-pseudoscalar-meson interactions in the SU(3) flavor basis at low energies
can be described by the simple chiral Lagrangian~\cite{Li:1997gd,Zhao:2002id}
\begin{equation}\label{coup}
H_{m}=\sum_j
\frac{1}{f_m}\bar{\psi}_j\gamma^{j}_{\mu}\gamma^{j}_{5}\psi_j\partial^{\mu}\phi_m,
\end{equation}
where $\psi_j$ represents the $j$th quark field in the hadron,
$\phi_m$ is the pseudoscalar meson field, and $f_m$ is the
pseudoscalar meson decay constant.
To match the nonrelativistic harmonic oscillator wave functions
adopted in the calculations, one should adopt the quark-pseudoscalar-meson interactions in
the nonrelativistic form~\cite{Li:1997gd,Zhao:2002id,Zhong:2008kd,Zhong:2010vq,Zhong:2009sk,
Zhong:2007gp,Liu:2012sj,Xiao:2013xi,Wang:2017hej,Xiao:2014ura}:
\begin{eqnarray}\label{ccpk}
H_{m}^{nr}=\sum_j\left[ \mathcal{G} \vsig_j \cdot \textbf{q}
+h \vsig_j\cdot \textbf{p}_j\right]I_j
e^{-i\mathbf{q}\cdot \mathbf{r}_j},
\end{eqnarray}
with $\mathcal{G}\equiv -(1+\frac{\omega_m}{E_f+M_f})$ and $h\equiv \frac{\omega_m}{2\mu_q}$.
In the above equation, $\vsig_j$ and $\textbf{p}_j$ stand for the
Pauli spin vector and internal momentum operator for the $j$th quark of the initial hadron;
$\omega_m$ and $\mathbf{q}$ stand for the energy and three momenta of the emitted light meson,
respectively; $\mu_q$ is a reduced mass given by $1/\mu_q=1/m_j+1/m'_j$ with
$m_j$ and $m'_j$ for the masses of the $j$th quark in the initial and
final hadrons, respectively; and $I_j$ is the isospin operator associated
with the pseudoscalar meson~\cite{Li:1997gd,Zhong:2008kd}
\begin{equation} I_j=\begin{cases}
        a^{\dagger}_j(u)a_j(s)   &$for$~K^-,\\
        a^{\dagger}_j(d)a_j(s)   &$for$~\bar{K}^0,\\
        a^{\dagger}_j(d)a_j(u)   &$for$~\pi^+,\\
        a^{\dagger}_j(u)a_j(d)   &$for$~\pi^-,\\
        \frac{1}{\sqrt{2}}[a^{\dagger}_j(u)a_j(u)-a^{\dagger}_j(d)a_j(d)]    &$for$~\pi^0.
       \end{cases}
\end{equation}
It should be mentioned that the nonrelativistic form
of quark-pseudoscalar-meson interactions expressed in Eq.~(\ref{ccpk}) is similar to that in Refs.~\cite{Godfrey:1985xj,Koniuk:1979vy,Capstick:2000qj},
except that the factors $\mathcal{G}$ and $h$ in this work have an explicit
dependence on the energies of final hadrons.

Meanwhile, to treat the radiative decay of a hadron we apply
the constituent quark model, which has been successfully applied to study
the radiative decays of $c\bar{c}$ and $b\bar{b}$ systems~\cite{Deng:2016stx,Deng:2016ktl}
and heavy baryons~\cite{Wang:2017hej,Xiao:2017udy}.
In this model, the quark-photon EM coupling at the tree level
is adopted as
\begin{eqnarray}\label{he}
H_e=-\sum_j
e_{j}\bar{\psi}_j\gamma^{j}_{\mu}A^{\mu}(\mathbf{k},\mathbf{r}_j)\psi_j,
\end{eqnarray}
where $A^{\mu}$ represents the photon field with three momenta $\mathbf{k}$. $e_j$ and $\mathbf{r}_j$
stand for the charge and coordinate of the constituent quark $\psi_j$, respectively.
Similarly, to match the nonrelativistic harmonic oscillator wave functions, we adopt
the nonrelativistic form of the quark-photon EM couplings~\cite{Deng:2016stx,Deng:2016ktl,Brodsky:1968ea,Li:1997gd,Zhao:1998fn,Zhao:2002id,Xiao:2015gra,Zhong:2011ti,Zhong:2011ht},
\begin{equation}\label{he2}
H_{e}^{nr}=\sum_{j}\left[e_{j}\mathbf{r}_{j}\cdot\veps-\frac{e_{j}}{2m_{j}
}\vsig_{j}\cdot(\veps\times\hat{\mathbf{k}})\right]e^{-i\textbf{k}\cdot
\textbf{r}_j},
\end{equation}
where $\boldsymbol \epsilon$ is the polarization vector of the final photon.
The first and second terms in Eq.(\ref{he2}) are responsible for the electric and
magnetic transitions, respectively. The second term
in Eq.(\ref{he2}) is the same as that used in
Refs.~\cite{Godfrey:1985xj,Koniuk:1979vy,Capstick:2000qj,Copley:1969ft,Sartor:1986sf}, while the first term in Eq.(\ref{he2})
differs from $(1/m_j) \mathbf{p}_j\cdot\veps$ used in Refs.~\cite{Godfrey:1985xj,Koniuk:1979vy,Capstick:2000qj,Copley:1969ft,Sartor:1986sf}
in order to take into account the binding potential effects~\cite{Brodsky:1968ea}.

For a light pseudoscalar meson emission in a strong decay process,
the partial decay width can be calculated with~\cite{Zhong:2008kd, Zhong:2007gp}
\begin{equation}\label{dww}
\Gamma_m=\left(\frac{\delta}{f_m}\right)^2\frac{(E_f+M_f)|\mathbf{q}|}{4\pi
M_i(2J_i+1)} \sum_{J_{fz},J_{iz}}|\mathcal{M}_{J_{fz},J_{iz}}|^2 ,
\end{equation}
while for a photon emission in a radiative decay process, the partial decay width can be calculated with~\cite{Deng:2016stx,Deng:2016ktl}
\begin{equation}\label{dww}
\Gamma_\gamma=\frac{|\mathbf{k}|^2}{\pi}\frac{2}{2J_i+1}\frac{M_{f}}{M_{i}}\sum_{J_{fz},J_{iz}}|\mathcal{A}_{J_{fz},J_{iz}}|^2,
\end{equation}
where $\mathcal{M}_{J_{fz},J_{iz}}$ and $\mathcal{A}_{J_{fz},J_{iz}}$ correspond to
the strong and radiative transition amplitudes, respectively.
The quantum numbers $J_{iz}$ and $J_{fz}$ stand for the third components of the total
angular momenta of the initial and final heavy baryons,
respectively. $\mathbf{q}$ stands for the three momenta
of the emitted pseudoscalar meson.
$E_f$ and $M_f$ are the energy and mass of the final heavy baryon, and
$M_i$ is the mass of the initial heavy baryon.
$\delta$ as a global parameter accounts for the
strength of the quark-meson couplings. It has been determined in our previous study of the strong
decays of the charmed baryons and heavy-light mesons
\cite{Zhong:2007gp,Zhong:2008kd}. Here, we fix its value the same as
that in Refs.~\cite{Zhong:2008kd,Zhong:2007gp}, i.e. $\delta=0.557$.

In the calculation, the standard quark model parameters are
adopted. Namely, we set $m_u=m_d=330$ MeV, $m_s = 450$ MeV, $m_c = 1480$
MeV and $m_b=5000$ MeV for the constituent quark masses.
Considering the mass differences between the $u/d$ and $s$ constituent quarks,
the harmonic oscillator parameter $\alpha_{\rho}$ in the wave function
$\psi^n_{lm}=R_{nl}Y_{lm}$ for $uu/ud/dd$, $us/ds$
and $ss$ diquark systems should be different from each other.
Thus, we take $\alpha_{\rho}=400$, 420 and $440$ MeV for $uu/ud/dd$, $us/ds$
and $ss$ diquark systems, respectively. Another harmonic oscillator parameter $\alpha_{\lambda}$ can
be related to $\alpha_{\rho}$ by the relation given in Eq.~(\ref{rhol}). The decay
constants for $\pi$ and $K$ mesons are taken as $f_\pi = 132$ MeV and
$f_K = 160$ MeV, respectively. The masses
of the well-established hadrons used in the calculations are
adopted from the PDG~\cite{Olive:2016xmw}.

\begin{table*}[htp]
\begin{center}
\caption{ \label{rad2} Partial decay widths (keV) for the radiative decays of the $1P$ singly
heavy baryons belonging to $\bar{\mathbf{3}}_F$. Ours (A) stands for the results predicted by assuming
the singly heavy baryons as the $\lambda$-mode excitations.  Ours (B,C) stand for the results predicted by assuming
the singly heavy baryons as the $\rho$-mode excitations with spin quantum numbers $S=1/2,3/2$, respectively. }
%\footnotesize
\begin{tabular}{ccccccccccccccccccccccccc}
\hline\hline
$B_Q \rightarrow B_Q^{'}$      &Ours (A) &Ours (B) & Ours (C) &RQM~\cite{Ivanov:1999bk}   &LCQSR~\cite{Zhu:2000py}
&HQS~\cite{Tawfiq:1999cf} &Bound state~\cite{Chow:1995nw}   &Bound state~\cite{Gamermann:2010ga}\\
\hline
$\Lambda_{c}(2593)\frac{1}{2}^- \rightarrow \Lambda_c^{+}\gamma$   &~~0.26~~   &1.59&~~~0.80   &$115\pm1$         & 36               &$\cdot\cdot\cdot$ & 16                &278   \\
$\Lambda_{c}(2593)\frac{1}{2}^- \rightarrow \Sigma_c^{+}\gamma$    &~~0.45~~   &41.6&~~~0.08   &$77\pm1$          &  11              &$\cdot\cdot\cdot$ &$\cdot\cdot\cdot$  &2     \\
$\Lambda_{c}(2593)\frac{1}{2}^- \rightarrow \Sigma_c^{*+}\gamma$   &~~0.05~~   &0.02&~~~6.81   &$6\pm1$           &  1               & 6.05             &$\cdot\cdot\cdot$  &$\cdot\cdot\cdot$\\
$\Lambda_{c}(2625)\frac{3}{2}^- \rightarrow \Lambda_c^{+}\gamma$   &~~0.30~~   &2.35&~~~3.29   &$151\pm2$         & 48               &$\cdot\cdot\cdot$ & 21                &$\cdot\cdot\cdot$\\
$\Lambda_{c}(2625)\frac{3}{2}^- \rightarrow \Sigma_c^{+}\gamma$    &~~1.17~~   &48.0&~~~0.55   &$35\pm0.5$        & 5                & 34.7             &$\cdot\cdot\cdot$  &$\cdot\cdot\cdot$\\
$\Lambda_{c}(2625)\frac{3}{2}^- \rightarrow \Sigma_c^{*+}\gamma$   &~~0.26~~   &0.09&~~~17.4   &$46\pm0.6$        & 6                & 43.2             &$\cdot\cdot\cdot$  &$\cdot\cdot\cdot$\\
$\Xi_{c}^{+}(2790)\frac{1}{2}^- \rightarrow \Xi_c^{+}\gamma$       &~~4.65~~   &1.39&~~~0.75   &$\cdot\cdot\cdot$ &$\cdot\cdot\cdot$ &$\cdot\cdot\cdot$ &$\cdot\cdot\cdot$  &   246 \\
$\Xi_{c}^{0}(2790)\frac{1}{2}^- \rightarrow \Xi_c^{0}\gamma$       &~~263~~~   &5.57&~~~3.00   &$\cdot\cdot\cdot$ &$\cdot\cdot\cdot$ &$\cdot\cdot\cdot$ &$\cdot\cdot\cdot$  &  117  \\
$\Xi_{c}^{+}(2790)\frac{1}{2}^- \rightarrow \Xi_c^{\prime+}\gamma$ &~~1.43~~   &128 &~~~0.41   &$\cdot\cdot\cdot$ &$\cdot\cdot\cdot$ &$\cdot\cdot\cdot$ &$\cdot\cdot\cdot$  &   1   \\
$\Xi_{c}^{0}(2790)\frac{1}{2}^- \rightarrow \Xi_c^{\prime0}\gamma$ &~~0.0~~~   &0.0 &~~~0.0    &$\cdot\cdot\cdot$ &$\cdot\cdot\cdot$ &$\cdot\cdot\cdot$ &$\cdot\cdot\cdot$  &   1   \\
$\Xi_{c}^{+}(2790)\frac{1}{2}^- \rightarrow \Xi_c^{*+}\gamma$      &~~0.44~~   &0.25&~~~43.4   &$\cdot\cdot\cdot$ &$\cdot\cdot\cdot$ &$\cdot\cdot\cdot$ &$\cdot\cdot\cdot$  &$\cdot\cdot\cdot$ \\
$\Xi_{c}^{0}(2790)\frac{1}{2}^- \rightarrow \Xi_c^{*0}\gamma$      &~~0.0~~~   &0.0 &~~~ 0.0    &$\cdot\cdot\cdot$ &$\cdot\cdot\cdot$ &$\cdot\cdot\cdot$ &$\cdot\cdot\cdot$  &$\cdot\cdot\cdot$ \\
$\Xi_{c}^{+}(2815)\frac{3}{2}^- \rightarrow \Xi_c^{+}\gamma$       &~~2.8~~~   &1.88&~~~2.81   &$190\pm5$         &$\cdot\cdot\cdot$ &$\cdot\cdot\cdot$ &$\cdot\cdot\cdot$  &$\cdot\cdot\cdot$ \\
$\Xi_{c}^{0}(2815)\frac{3}{2}^- \rightarrow \Xi_c^{0}\gamma$       &~~292~~~   &7.50&~~~11.2   &$497\pm14$        &$\cdot\cdot\cdot$ &$\cdot\cdot\cdot$ &$\cdot\cdot\cdot$  &$\cdot\cdot\cdot$ \\
$\Xi_{c}^{+}(2815)\frac{3}{2}^- \rightarrow \Xi_c^{\prime+}\gamma$ &~~2.32~~   &110 &~~~1.85   &$\cdot\cdot\cdot$ &$\cdot\cdot\cdot$ &$\cdot\cdot\cdot$ &$\cdot\cdot\cdot$  &$\cdot\cdot\cdot$ \\
$\Xi_{c}^{0}(2815)\frac{3}{2}^- \rightarrow \Xi_c^{\prime0}\gamma$ &~~0.0~~~   &0.0 &~~~ 0.0    &$\cdot\cdot\cdot$ &$\cdot\cdot\cdot$ &$\cdot\cdot\cdot$ &$\cdot\cdot\cdot$  &$\cdot\cdot\cdot$ \\
$\Xi_{c}^{+}(2815)\frac{3}{2}^- \rightarrow \Xi_c^{*+}\gamma$      &~~0.99~~   &0.52&~~~58.1   &$\cdot\cdot\cdot$ &$\cdot\cdot\cdot$ &$\cdot\cdot\cdot$ &$\cdot\cdot\cdot$  &$\cdot\cdot\cdot$ \\
$\Xi_{c}^{0}(2815)\frac{3}{2}^- \rightarrow \Xi_c^{*0}\gamma$      &~~0.0~~~   &0.0 &~~~0.0    &$\cdot\cdot\cdot$ &$\cdot\cdot\cdot$ &$\cdot\cdot\cdot$ &$\cdot\cdot\cdot$  &$\cdot\cdot\cdot$ \\
$\Lambda_{b}(5912)\frac{1}{2}^- \rightarrow \Lambda_b^{0}\gamma$   &~~50.2~~   &1.62&~~~0.81   &$\cdot\cdot\cdot$ & 1                &$\cdot\cdot\cdot$ & 90                &$\cdot\cdot\cdot$ \\
$\Lambda_{b}(5912)\frac{1}{2}^- \rightarrow \Sigma_b^{0}\gamma$    &~~0.14~~   &16.2&~~~0.02   &$\cdot\cdot\cdot$ & 11               &81.7              &$\cdot\cdot\cdot$  &$\cdot\cdot\cdot$\\
$\Lambda_{b}(5912)\frac{1}{2}^- \rightarrow \Sigma_b^{*0}\gamma$   &~~0.09~~   &0.02&~~~8.25   &$\cdot\cdot\cdot$ & 1                &8.91              &$\cdot\cdot\cdot$  &$\cdot\cdot\cdot$ \\
$\Lambda_{b}(5920)\frac{3}{2}^- \rightarrow \Lambda_b^{0}\gamma$   &~~52.8~~   &1.81&~~~2.54   &$\cdot\cdot\cdot$ &  1               &$\cdot\cdot\cdot$ &119                &$\cdot\cdot\cdot$ \\
$\Lambda_{b}(5920)\frac{3}{2}^- \rightarrow \Sigma_b^{0}\gamma$    &~~0.21~~   &15.1&~~~0.07   &$\cdot\cdot\cdot$ &  5               &33.8              &$\cdot\cdot\cdot$  &$\cdot\cdot\cdot$\\
$\Lambda_{b}(5920)\frac{3}{2}^- \rightarrow \Sigma_b^{*0}\gamma$   &~~0.15~~   &0.03&~~~9.90   &$\cdot\cdot\cdot$ &  6               &49.9              &$\cdot\cdot\cdot$  &$\cdot\cdot\cdot$\\
$\Xi_{b}^{0}(6120)\frac{1}{2}^- \rightarrow \Xi_b^{0}\gamma$       &~~63.6~~   &1.86&~~~0.93   &$\cdot\cdot\cdot$ &$\cdot\cdot\cdot$ &$\cdot\cdot\cdot$ &$\cdot\cdot\cdot$  &$\cdot\cdot\cdot$\\
$\Xi_{b}^{-}(6120)\frac{1}{2}^- \rightarrow \Xi_b^{-}\gamma$       &~~135~~~   &7.19&~~~3.59   &$\cdot\cdot\cdot$ &$\cdot\cdot\cdot$ &$\cdot\cdot\cdot$ &$\cdot\cdot\cdot$  &$\cdot\cdot\cdot$ \\
$\Xi_{b}^{0}(6120)\frac{1}{2}^- \rightarrow \Xi_b^{\prime0}\gamma$ &~~1.32~~   &94.3&~~~0.16   &$\cdot\cdot\cdot$ &$\cdot\cdot\cdot$ &$\cdot\cdot\cdot$ &$\cdot\cdot\cdot$  &$\cdot\cdot\cdot$\\
$\Xi_{b}^{-}(6120)\frac{1}{2}^- \rightarrow \Xi_b^{\prime-}\gamma$ &~~0.0~~~   &0.0 &~~~ 0.0    &$\cdot\cdot\cdot$ &$\cdot\cdot\cdot$ &$\cdot\cdot\cdot$ &$\cdot\cdot\cdot$  &$\cdot\cdot\cdot$ \\
$\Xi_{b}^{0}(6120)\frac{1}{2}^- \rightarrow \Xi_b^{*0}\gamma$      &~~2.04~~   &0.62&~~~80.0   &$\cdot\cdot\cdot$ &$\cdot\cdot\cdot$ &$\cdot\cdot\cdot$ &$\cdot\cdot\cdot$  &$\cdot\cdot\cdot$\\
$\Xi_{b}^{-}(6120)\frac{1}{2}^- \rightarrow \Xi_b^{*-}\gamma$      &~~0.0~~~   &0.0 &~~~ 0.0    &$\cdot\cdot\cdot$ &$\cdot\cdot\cdot$ &$\cdot\cdot\cdot$ &$\cdot\cdot\cdot$  &$\cdot\cdot\cdot$ \\
$\Xi_{b}^{0}(6130)\frac{3}{2}^- \rightarrow \Xi_b^{0}\gamma$       &~~68.3~~   &2.10&~~~2.94   &$\cdot\cdot\cdot$ &$\cdot\cdot\cdot$ &$\cdot\cdot\cdot$ &$\cdot\cdot\cdot$  &$\cdot\cdot\cdot$\\
$\Xi_{b}^{-}(6130)\frac{3}{2}^- \rightarrow \Xi_b^{-}\gamma$       &~~147~~~   &8.13&~~~11.4   &$\cdot\cdot\cdot$ &$\cdot\cdot\cdot$ &$\cdot\cdot\cdot$ &$\cdot\cdot\cdot$  &$\cdot\cdot\cdot$ \\
$\Xi_{b}^{0}(6130)\frac{3}{2}^- \rightarrow \Xi_b^{\prime0}\gamma$ &~~1.68~~   &69.4&~~~0.80   &$\cdot\cdot\cdot$ &$\cdot\cdot\cdot$ &$\cdot\cdot\cdot$ $\cdot\cdot\cdot$ &&$\cdot\cdot\cdot$\\
$\Xi_{b}^{-}(6130)\frac{3}{2}^- \rightarrow \Xi_b^{\prime-}\gamma$ &~~0.0~~~   &0.0 & ~~~ 0.0    &$\cdot\cdot\cdot$ &$\cdot\cdot\cdot$ &$\cdot\cdot\cdot$ &$\cdot\cdot\cdot$ &$\cdot\cdot\cdot$ \\
$\Xi_{b}^{0}(6130)\frac{3}{2}^- \rightarrow \Xi_b^{*0}\gamma$      &~~2.64~~   &0.80&~~~78.0   &$\cdot\cdot\cdot$ &$\cdot\cdot\cdot$ &$\cdot\cdot\cdot$ &$\cdot\cdot\cdot$ &$\cdot\cdot\cdot$\\
$\Xi_{b}^{-}(6130)\frac{3}{2}^- \rightarrow \Xi_b^{*-}\gamma$      &~~0.0~~~   &0.0 & ~~~ 0.0    &$\cdot\cdot\cdot$ &$\cdot\cdot\cdot$ &$\cdot\cdot\cdot$ &$\cdot\cdot\cdot$ &$\cdot\cdot\cdot$ \\
\hline\hline
\end{tabular}
\end{center}
\end{table*}

\begin{table*}[htp]
\begin{center}
\caption{ \label{Rad} Partial decay widths (keV) for the radiative decays of the $1S$-wave singly heavy baryons belonging to $\mathbf{6}_F$.}
%\footnotesize
\begin{tabular}{p{2.0cm}p{1.2cm}p{1.8cm}p{1.8cm}p{1.6cm}p{2.2cm}p{1.5cm}p{1.7cm}p{2.7cm} c|c|c|c|c|}
%|p{2.0cm}|p{1.5cm}|p{1.5cm}|p{1.5cm}|p{1.5cm}|p{1.8cm}|p{1.3cm}|p{1.3cm}|p{1.5cm}
\hline\hline
$B_Q \rightarrow B_Q^{'}$                 &Ours    &RQM~\cite{Ivanov:1999bk}   &VMC~\cite{Aliev:2011bm}& BM~\cite{Bernotas:2013eia}  &LCQSR~\cite{Aliev:2009jt,Aliev:2016xvq,Aliev:2014bma}
&HQS~\cite{Tawfiq:1999cf}   &NQM~\cite{Dey:1994qi} &Other works  \\ \hline
$\Sigma_c^{+} \rightarrow \Lambda_c^{+}\gamma$   &80.6     &$60.7\pm1.5$        &$\cdot\cdot\cdot$ &46.1   &$50\pm 17$       &$\cdot\cdot\cdot$  &98.7                &93~\cite{Cheng:1992xi}  \\
$\Sigma_c^{*+} \rightarrow \Lambda_c^{+}\gamma$  &373      &$151\pm4$           &409.3             &126    &$130\pm65$       &233                &250                 &$\cdot\cdot\cdot$       \\
$\Sigma_c^{*+} \rightarrow \Sigma_c^{+}\gamma$   &0.004    &$0.14\pm0.004$      &0.187             &0.004  &$0.40\pm0.22$    &0.22               &$1\times10^{-3}$    &$0.40^{+0.43}_{-0.21}$\cite{Wang:2009ic}  \\
$\Sigma_c^{*0} \rightarrow \Sigma_c^{0}\gamma$   &3.43     &$\cdot\cdot\cdot$   &1.049             &1.08   &$0.08\pm0.042$   &$\cdot\cdot\cdot$  &1.2                 &$1.58^{+1.68}_{-0.82}$~\cite{Wang:2009ic}  \\
$\Sigma_c^{*++} \rightarrow \Sigma_c^{++}\gamma$ &3.94     &$\cdot\cdot\cdot$   &3.567             &0.826  &$2.65\pm1.60$    &$\cdot\cdot\cdot$  &1.7                 &$6.36^{+6.79}_{-3.31}$~\cite{Wang:2009ic}  \\
$\Xi_c^{\prime+} \rightarrow \Xi_c^{+}\gamma$    &42.3     &$12.7\pm1.5$        &$\cdot\cdot\cdot$ &10.2   &$8.5\pm2.5$      &14.6               &32                  &16~\cite{Cheng:1992xi}, 5.468~\cite{Bahtiyar:2016dom}\\
$\Xi_c^{\prime0} \rightarrow \Xi_c^{0}\gamma$    &0.0      &$0.17\pm0.02$       &$\cdot\cdot\cdot$ &0.0015 &$0.27\pm 0.06$   &0.35               &0.27                &0.3~\cite{Cheng:1992xi}, 0.002~\cite{Bahtiyar:2016dom} \\
$\Xi_c^{\prime*+} \rightarrow \Xi_c^{+}\gamma$         &139      &$54\pm3$            & 152.4            &44.3   &$52\pm32$        &$\cdot\cdot\cdot$  &124                 &$\cdot\cdot\cdot$       \\
$\Xi_c^{\prime*0} \rightarrow \Xi_c^{0}\gamma$         &0.0      &$0.68\pm0.04$       &1.318             &0.908  &$0.66\pm0.41$    &$\cdot\cdot\cdot$  &0.8                 &$\cdot\cdot\cdot$       \\
$\Xi_c^{\prime*+} \rightarrow \Xi_c^{\prime+}\gamma$   &$0.004$  &$\cdot\cdot\cdot$   &0.485             &0.011  &$0.274$          &$\cdot\cdot\cdot$  &0.03                &$\cdot\cdot\cdot$       \\
$\Xi_c^{*0} \rightarrow \Xi_c^{\prime0}\gamma$   &3.03     &$\cdot\cdot\cdot$   &1.317             &1.03   &$2.142$          &$\cdot\cdot\cdot$  &0.7                 &$\cdot\cdot\cdot$       \\
$\Omega_c^{*0} \rightarrow \Omega_c^{0}\gamma$   &0.89     &$\cdot\cdot\cdot$   &1.439             &1.07   &$0.932$          &$\cdot\cdot\cdot$  &$\cdot\cdot\cdot$   & $ 0.074(8)$~\cite{Bahtiyar:2015sga}\\
$\Sigma_b^{0} \rightarrow \Lambda_b^{0}\gamma$   &130      &$\cdot\cdot\cdot$   &$\cdot\cdot\cdot$ &58.9   &$152\pm 60$      &$\cdot\cdot\cdot$  &$\cdot\cdot\cdot$   &$\cdot\cdot\cdot$       \\
$\Sigma_b^{*0} \rightarrow \Lambda_b^{0}\gamma$  &335      &$\cdot\cdot\cdot$   &221.5             &81.1   &$114\pm62$       &251                &$\cdot\cdot\cdot$   &344~\cite{Zhu:1998ih}    \\
$\Sigma_b^{*0} \rightarrow \Sigma_b^{0}\gamma$   &0.02     &$\cdot\cdot\cdot$   &0.006             &0.005  &$0.028\pm0.020$  &0.15               &$\cdot\cdot\cdot$   &0.08~\cite{Zhu:1998ih}   \\
$\Sigma_b^{*+} \rightarrow \Sigma_b^{+}\gamma$   &0.25     &$\cdot\cdot\cdot$   &0.137             &0.054  &$0.46\pm0.28$    &$\cdot\cdot\cdot$  &$\cdot\cdot\cdot$   &1.26~\cite{Zhu:1998ih}   \\
$\Sigma_b^{*-} \rightarrow \Sigma_b^{-}\gamma$   &0.06     &$\cdot\cdot\cdot$   &0.040             &0.01   &$0.11\pm0.076$   &$\cdot\cdot\cdot$  &$\cdot\cdot\cdot$   &0.32~\cite{Zhu:1998ih}   \\
$\Xi_b^{\prime0} \rightarrow \Xi_b^{0}\gamma$    &84.6     &$\cdot\cdot\cdot$   &$\cdot\cdot\cdot$ &14.7   &$47\pm 21$       &$\cdot\cdot\cdot$  &$\cdot\cdot\cdot$   &$\cdot\cdot\cdot$       \\
$\Xi_b^{\prime-} \rightarrow \Xi_b^{-}\gamma$    &0.0      &$\cdot\cdot\cdot$   &$\cdot\cdot\cdot$ &0.118  &$3.3\pm 1.3$     &$\cdot\cdot\cdot$  &$\cdot\cdot\cdot$   &$\cdot\cdot\cdot$       \\
$\Xi_b^{\prime*0} \rightarrow \Xi_b^{0}\gamma$         &104      &$\cdot\cdot\cdot$   &270.8             &24.7   &$135\pm85$       &$\cdot\cdot\cdot$  &$\cdot\cdot\cdot$   &$\cdot\cdot\cdot$       \\
$\Xi_b^{\prime*-} \rightarrow \Xi_b^{-}\gamma$         &0.0      &$\cdot\cdot\cdot$   &2.246             &0.278  &$1.50\pm0.95$    &$\cdot\cdot\cdot$  &$\cdot\cdot\cdot$   &$\cdot\cdot\cdot$       \\
$\Xi_b^{\prime*0} \rightarrow \Xi_b^{\prime0}\gamma$   &5.19     &$\cdot\cdot\cdot$   &0.281             &0.004  &$0.131$          &$\cdot\cdot\cdot$  &$\cdot\cdot\cdot$   &$\cdot\cdot\cdot$       \\
$\Xi_b^{\prime*-} \rightarrow \Xi_b^{\prime-}\gamma$   &15.0     &$\cdot\cdot\cdot$   &0.702             &0.005  &$0.303$          &$\cdot\cdot\cdot$  &$\cdot\cdot\cdot$   &$\cdot\cdot\cdot$       \\
$\Omega_b^{*-} \rightarrow \Omega_b^{-}\gamma$   &0.1      &$\cdot\cdot\cdot$   &2.873             & 0.006 &$0.092$          &$\cdot\cdot\cdot$  &$\cdot\cdot\cdot$   &$\cdot\cdot\cdot$       \\
\hline\hline
\end{tabular}
\end{center}
\end{table*}

\section{Results and discussions}\label{results}

\subsection{The $\Lambda_c$ and $\Lambda_b$ states }\label{LLL}

%\subsubsection{$1P$-wave $\Lambda_c$ states }

In the $\Lambda_c$ and $\Lambda_b$ families,
there are two $\lambda$-mode $1P$ excitations $|^2P_\lambda \frac{1}{2}^-\rangle$
and $|^2P_\lambda \frac{3}{2}^-\rangle$ according to the quark model classification.
$\Lambda_c(2593)$ and $\Lambda_c(2625)$ could be assigned to the
$\lambda$-mode $1P$ excitations with $J^P=1/2^-$ and $3/2^-$, respectively.
With these assignments, the strong decays of both $\Lambda_c(2593)$ and
$\Lambda_c(2625)$ into $\Sigma_c\pi$ can be well understood within the ChQM~\cite{Zhong:2007gp}.
In this work, assuming $\Lambda_c(2593)$ and $\Lambda_c(2625)$ as
the $\lambda$-mode $1P$ excitations, we further estimate their radiative decays into $\Lambda_c\gamma$,
$\Sigma_c\gamma$ and $\Sigma_c^{*}\gamma$ within the constituent quark model.
Our results have been listed in Table~\ref{Rad2} (denoted by Ours (A)).

It shows that the expected partial radiative decay widths of $\Gamma [\Lambda_c(2593)\to \Lambda_c^+\gamma, \Sigma_c^+\gamma]$
and $\Gamma [\Lambda_c(2625)^+\to \Lambda_c^+\gamma, \Sigma_c^+\gamma,
\Sigma_c^{*+}\gamma]$ are at the order of $\sim 200- 400$ eV. Combining the predicted width
$\Gamma [\Lambda_c(2625)]\simeq 50$ keV for $\Lambda_c(2625)^+$~\cite{Zhong:2007gp},
we estimate that the branching fractions of $\mathcal{B} [\Lambda_c(2625)^+\to \Lambda_c^+\gamma, \Sigma_c^+\gamma,
\Sigma_c^{*+}\gamma]$ may reach up to $O(10^{-3})$.
The relatively large branching fractions indicate that
the radiative decay processes, $\Lambda_c(2625)^+\to \Lambda_c^+\gamma, \Sigma_c^+\gamma,
\Sigma_c^{*+}\gamma$, can be accessible in future experiments.

While combining the measured width $\Gamma [\Lambda_c(2593)]\simeq 2.59$ MeV
for $\Lambda_c(2593)$~\cite{Olive:2016xmw} with our predicted partial widths, the branching fractions of
$\mathcal{B}[\Lambda_c(2593)^+\to \Lambda_c^+\gamma, \Sigma_c^+\gamma]$
are estimated to be $O(10^{-5})$. The tiny branching fractions indicate that
measurements of the radiative decays of $\Lambda_c(2593)$
in future experiments could be more difficult.

As a comparison, the predictions from other models
are also listed in Table~\ref{Rad2}. It shows that our results for
$\Gamma [\Lambda_c(2593)^+\to \Lambda_c^+\gamma, \Sigma_c^+\gamma]$
and $\Gamma [\Lambda_c(2625)^+\to \Lambda_c^+\gamma, \Sigma_c^+\gamma,
\Sigma_c^{*+}\gamma]$ are about 2-3 orders of magnitude smaller than
the predictions from RQM~\cite{Ivanov:1999bk}.
The results from LCQSR~\cite{Zhu:2000py} are
about an order of magnitude smaller than those of
Ref.~\cite{Ivanov:1999bk}, but still much larger than our results. It should be mentioned that within the
meson-baryon bound state picture, the radiative decays
of $\Lambda_c(2593)$ were studied by two groups as well~\cite{Gamermann:2010ga,Chow:1995nw}. However,
their results for $\Gamma[\Lambda_c(2593)^+\to \Lambda_c^+\gamma]$ turn out to have significant discrepancies (see Table~\ref{Rad2}).

To further understand the small radiative partial widths that we obtain in the ChQM, we further analyze
the transition amplitudes. The extracted amplitude for
the $\Lambda_c(2593)^+\to \Lambda_c^+\gamma$ transition is expressed as
\begin{eqnarray}\label{AAA}
i\mathcal{A}_{\frac{1}{2},-\frac{1}{2}}&=&\frac{\langle e_1 \rangle }{\alpha_\lambda}\frac{1}{2} \frac{\sqrt{k} m'}{2m+m'}\exp\left(-\frac{k_\rho^{2}}{4\alpha_\rho^{2}}-\frac{k_\lambda^{2}}{4\alpha_\lambda^{2}}\right)\nonumber \\
&& +\frac{\langle e_2 \rangle }{\alpha_\lambda}\frac{1}{2} \frac{\sqrt{k} m'}{2m+m'}\exp\left(-\frac{k_\rho^{2}}{4\alpha_\rho^{2}}-\frac{k_\lambda^{2}}{4\alpha_\lambda^{2}}\right) \nonumber \\
&&- \frac{\langle e_3 \rangle }{\alpha_\lambda}\left( \frac{\sqrt{k} m}{2m+m'}+\frac{\sqrt{k}k'_\lambda}{2\sqrt{6}m'}\right)\exp\left(-\frac{k_\lambda^{'2}}{4\alpha_\lambda^{2}}\right),
\end{eqnarray}
with $k_\rho=\sqrt{\frac{1}{2}}k$, $k_\lambda=\sqrt{\frac{1}{6}}\frac{3m'}{2m+m'}k$
and $k_\lambda'=\sqrt{\frac{2}{3}}\frac{3m}{2m+m'}k$. In Eq.~(\ref{AAA}),
the first two terms correspond to the contributions from the two light quarks $q_1$
and $q_2$ via an $E1$ transition, respectively. The third term stands for the
contributions from the heavy quark $Q$ (i.e., $c$ quark) via $E1$
and $M2$ transitions. Note that the matrix elements for the charge operators,
$\langle e_1 \rangle=\langle e_2 \rangle=1/6$ and $\langle e_3 \rangle=2/3$,
and the quark model form factor $\exp(-k_\rho^{2}/4\alpha_\rho^{2}-k_\lambda^{2}/4\alpha_\lambda^{2})\simeq
\exp(-k_\lambda^{'2}/4\alpha_\lambda^{2})\simeq 1$ at a low photon momentum $k$.
Using these conditions, one finds that the contributions from the two
light quarks $q_1$ and $q_2$ have a strong destructive interference with
the contributions from the heavy quark $Q$, which leads to a very small
amplitude $\mathcal{A}_{\frac{1}{2},-\frac{1}{2}}$ for $\Lambda_c(2593)^+\to \Lambda_c^+\gamma$.
A similar mechanism also exists in the transition amplitudes $\mathcal{A}_{\frac{1}{2},-\frac{1}{2}}$
and $\mathcal{A}_{-\frac{1}{2},-\frac{3}{2}}$ for $\Lambda_c(2625)^+\to \Lambda_c^+\gamma$.
This accounts for the small partial decay widths for $\Lambda_c(2593)^+\to \Lambda_c^+\gamma$
and $\Lambda_c(2625)^+\to \Lambda_c^+\gamma$.

In the $\Lambda_b$ family, the two $\lambda$-mode $1P$
excitations with $J^P=1/2^-$ and $3/2^-$ may correspond to
the newly observed states $\Lambda_b(5912)^0$ and $\Lambda_b(5920)^0$
at LHCb~\cite{Aaij:2012da}, respectively.
Their decays should be governed by the one-photon radiative
transitions because the strong decay channel $\Sigma_b\pi$ is not open.
Our estimate of their radiative decays are listed in Table \ref{Rad}.
It shows that the radiative decays of both states are dominated by the $\Lambda_b\gamma$ channel. The estimated partial decay widths are
\begin{eqnarray}\label{1DL}
\Gamma[\Lambda_b(5912)\to \Lambda_b\gamma]\simeq 50 \ \mathrm{keV}, \\
\Gamma[\Lambda_b(5920)\to \Lambda_b\gamma]\simeq 53 \ \mathrm{keV}.
\end{eqnarray}
Combining them with the measured widths $\Gamma[\Lambda_b(5912)^0]<0.66$ MeV
and $\Gamma[\Lambda_b(5920)^0]<0.63$ MeV~\cite{Aaij:2012da}, we obtain
$\mathcal{B}[\Lambda_b(5912,5920)\to \Lambda_b\gamma]>7\%$.
The large branching fractions indicate that
there are large potentials to observe
$\Lambda_b(5912,5920)\to \Lambda_b\gamma$ in future experiments.

In our model the decay rates of $\Lambda_b(5912)^0$ and $\Lambda_b(5920)^0$ into
the $\Lambda_b\gamma$ channel are predicted to be rather large which can be understood
by the decay amplitudes. The radiative decay amplitude of
$\Lambda_b(5912)^0\to \Lambda_b\gamma$ has the same form as
listed in Eq.~(\ref{AAA}). The only difference comes from
the matrix element of the charge operator of the heavy quark
$\langle e_3 \rangle$. For $\Lambda_b(5912)^0\to \Lambda_b\gamma$,
we have $\langle e_3 \rangle=-1/3$, which has a sign difference from
$\langle e_3 \rangle=2/3$ for $\Lambda_c(2593)\to \Lambda_c^+\gamma$. Thus,
the contributions from the two light quarks $q_1$ and $q_2$
have a constructive interference with the contributions from the heavy $b$ quark,
which leads to a large amplitude for
$\Lambda_b(5912)^0\to \Lambda_b\gamma$. Similar mechanism also exists in
the the transition amplitudes $\mathcal{A}_{\frac{1}{2},-\frac{1}{2}}$
and $\mathcal{A}_{-\frac{1}{2},-\frac{3}{2}}$ for $\Lambda_b(5920)^0\to \Lambda_b\gamma$ process.
Note that  our predictions of $\Gamma[\Lambda_b(5912,5920)\to \Lambda_b\gamma]$
are significantly larger than the predictions within LCQSR~\cite{Zhu:2000py}.

Finally, it should be mentioned that
$\Lambda_c(2593)$, $\Lambda_c(2625)$, $\Lambda_b(5912)^0$ and $\Lambda_b(5920)^0$
are suggested to be the $\rho$-mode excitations in Refs.~\cite{Chen:2015kpa,Mao:2015gya}.
Considering $\Lambda_c(2593)$, $\Lambda_c(2625)$,
$\Lambda_b(5912)^0$ and $\Lambda_b(5920)^0$ as the $P$-wave $\rho$-mode
excitations, we also study their radiative decays
within the constituent quark model. Our results are listed
in Table~\ref{rad2} as well. It shows that
most of the $\rho$-mode excitations in the radiative transitions are
different from the $\lambda$ mode.
To clarify the internal structure of $\Lambda_c(2593,2625)$ and $\Lambda_b(5912,5920)^0$,
their radiative decays are worth measuring in future
experiments.

\begin{figure*}[ht]
\centering \epsfxsize=8.2 cm \epsfbox{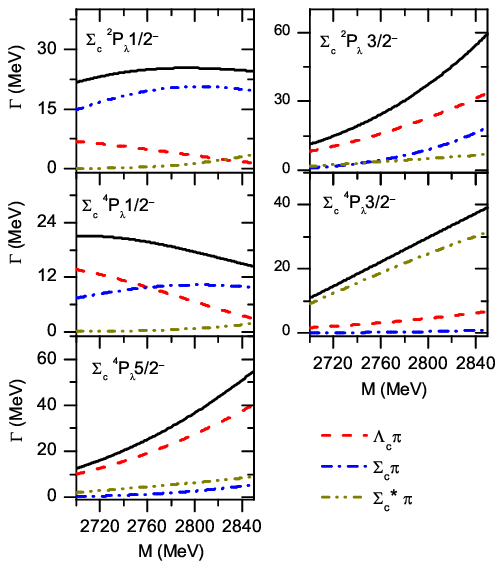} \epsfxsize=8.2 cm \epsfbox{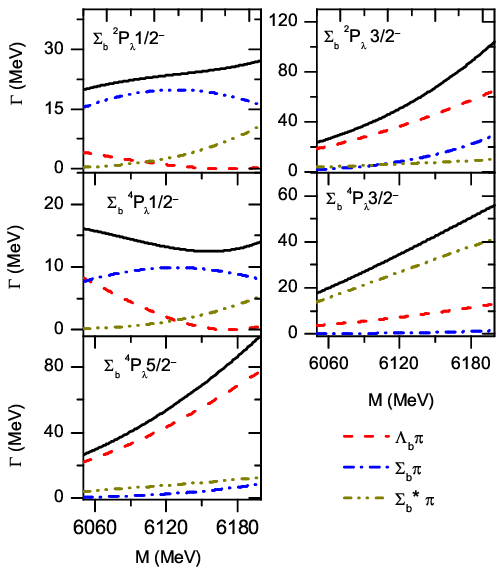} \vspace{-0.5cm} \caption{Partial and total strong decay widths of the $1P$ states in the $\Sigma_c$ and $\Sigma_b$
families as functions of their mass. The solid curves stand for the total widths.}\label{1pscb}
\end{figure*}

\begin{table*}[htb]
\begin{center}
\caption{ \label{sigemac} Partial widths (MeV) and branching fractions for the strong decays of the $1P$-wave states in the $\Sigma_c$ and $\Sigma_b$ families.}
%\footnotesize
\begin{tabular}{p{1.8cm}p{1.8cm}p{1.7cm}p{1.6cm}p{1.6cm}p{1.8cm}p{1.7cm}p{1.6cm}p{1.6cm}cccccccccccc}
\hline\hline
                  $|^{2S+1}L_\lambda~J^P\rangle$  & $\Sigma_c$ states  & Channel         &$\Gamma_{i}($MeV$)$          &$\mathcal{B}_i$   &$\Sigma_b$ states& Channel &$\Gamma_{th}($MeV$)$&$\mathcal{B}_i$     \\
\hline
$|^2P_{\lambda} \frac{1}{2}^- \rangle$  &$\Sigma_c(2713)$  &$\Lambda_c\pi$  &6.49  &28.65\% &$\Sigma_b(6101)$&$\Lambda_b\pi$ &1.74  &7.68\%  \\
&                                                          &$\Sigma_c\pi$   &16.08 &70.99\% &                &$\Sigma_b\pi$  &19.26 &85.00\% \\
&                                                          &$\Sigma_c^*\pi$ &0.08  &0.35\%  &                &$\Sigma_b^*\pi$&1.66  &7.33\%\\
&                                                          &total           &22.65 &        &                &total          &22.66 & \\
\hline
$|^2P_{\lambda} \frac{3}{2}^- \rangle$  &$\Sigma_c(2798)$  &$\Lambda_c\pi$  &22.53 &61.73\% &$\Sigma_b(6096)$&$\Lambda_b\pi$ &29.31 &74.60\% \\
&                                                          &$\Sigma_c\pi$   &8.84  &24.22\% &                &$\Sigma_b\pi$  &4.81  &12.24\% \\
&                                                          &$\Sigma_c^*\pi$ &5.13  &14.05\% &                &$\Sigma_b^*\pi$&5.17  &13.16\% \\
&                                                          &total           &36.5  &        &                &total          &39.29 &\\
\hline
$|^4P_{\lambda}\frac{1}{2}^- \rangle$  &$\Sigma_c(2799)$   &$\Lambda_c\pi$  &6.66  &37.78\% &$\Sigma_b(6095)$&$\Lambda_b\pi$ &4.00  &28.15\% \\
&                                                          &$\Sigma_c\pi$   &10.30 &58.42\% &                &$\Sigma_b\pi$  &9.50  &66.85\% \\
&                                                          &$\Sigma_c^*\pi$ &0.67  &3.80\%  &                &$\Sigma_b^*\pi$&0.71  &5.0\%  \\
&                                                          &total           &17.63 &        &                &total          &14.21 &\\
\hline
$|^4P_{\lambda} \frac{3}{2}^- \rangle$ &$\Sigma_c(2773)$   &$\Lambda_c\pi$  &3.62   &14.66\%&$\Sigma_b(6087)$&$\Lambda_b\pi$ &5.38   &20.46\%\\
&                                                         &$\Sigma_c\pi$    &0.29   &1.17\% &                &$\Sigma_b\pi$  &0.20   &0.76\%  \\
&                                                         &$\Sigma_c^*\pi$  &20.78  &84.16\%&                &$\Sigma_b^*\pi$&20.71  &78.78\%\\
&                                                         &total            &24.69  &&                    &total             &26.29  & \\
\hline
$|^4P_{\lambda} \frac{5}{2}^- \rangle$ &$\Sigma_c(2789)$  &$\Lambda_c\pi$   &25.03  &75.35\% &$\Sigma_b(6084)$&$\Lambda_b\pi$&31.38 &81.85\% \\
&                                                         &$\Sigma_c\pi$    &2.29   &6.89\%  &                &$\Sigma_b\pi$ &1.09  &2.84\%  \\
&                                                         &$\Sigma_c^*\pi$  &5.90   &17.76\% &                &$\Sigma_b^*\pi$ &5.77&15.05\% \\
&                                                         &total            &33.22  &        &                &total           &38.34  &\\
\hline
\end{tabular}
\end{center}
\end{table*}

\subsection{The $\Sigma_c$ and $\Sigma_b$ states }

\subsubsection{$1S$ states}

In the $\Sigma_c$ family, the assignment of $\Sigma_c(2455)$ and $\Sigma_c(2520)$ (i.e., $\Sigma_c$ and $\Sigma_c^*$) as
the $1S$ ground states with $J^P=1/2^+$ and $3/2^+$, respectively, is well accepted. In Ref.~\cite{Zhong:2007gp},
we have studied their strong decays into the $\Lambda_c\pi$ channel within the ChQM, our predictions were in good agreement with the data.
To better understand the properties of $\Sigma_c(2455)$ and $\Sigma_c(2520)$, we further study their radiative
decays in this work, and the results are listed in Table~\ref{Rad}. It shows that the singly charged states $\Sigma_c(2455)^+$ and $\Sigma_c(2520)^+$ have rather large radiative decay rates into the $\Lambda_c^+\gamma$ channel.
Their partial widths are predicted to be
\begin{eqnarray}\label{1ssc}
\Gamma[\Sigma_c(2455)^+\to \Lambda_c^+\gamma]&\simeq & 81 \ \mathrm{keV}, \\
\Gamma[\Sigma_c(2520)^+\to \Lambda_c^+\gamma]&\simeq & 373 \ \mathrm{keV},
\end{eqnarray}
which are consistent with other model predictions in magnitude (see Table~\ref{Rad}).
Combining our predicted partial widths with the average widths of
$\Sigma_c(2455)^+$ and $\Sigma_c(2520)^+$ from the PDG~\cite{Olive:2016xmw}, we estimate that the
branching fractions of $\mathcal{B}[\Sigma_c(2455,2520)^+\to \Lambda_c^+\gamma]$
can reach up to $2\%$. Thus, the radiative decay mode of
$\Sigma_c(2455,2520)^+\to \Lambda_c^+\gamma$ may be accessible in future experiments.

In the $\Sigma_b$ family, the $\Sigma_b^{\pm}$ and
$\Sigma_b^{*\pm}$ listed in the PDG book~\cite{Olive:2016xmw}
are naturally assigned to the $1S$ states with $J^P$ values $J^P=1/2^+$ and $3/2^+$,
respectively. It should be pointed out that
the neutral $1S$ states $\Sigma_b^{0}$ and
$\Sigma_b^{*0}$ are still missing.
We have studied the strong decays of $\Sigma_b^{(*)}$ within the ChQM in Ref.~\cite{Zhong:2007gp},
where their strong decays into $\Lambda_b\pi$ can be reasonably understood.
In this work, we further study their radiative decay properties and the results are listed in Table~\ref{Rad}.
It shows that the neutral states $\Sigma_b^{0}$ and $\Sigma_b^{*0}$ have
large radiative decay rates into $\Lambda_b^0\gamma$, and the partial widths
are predicted to be
\begin{eqnarray}\label{1ssb}
\Gamma[\Sigma_b^0\to \Lambda_b^0\gamma]\simeq 130 \ \mathrm{keV}, \\
\Gamma[\Sigma_b^{*0}\to \Lambda_b^0\gamma]\simeq 335 \ \mathrm{keV},
\end{eqnarray}
which are consistent with the other model predictions in magnitude (see Table~\ref{Rad}).

Combining these partial widths with the total widths, $\Gamma (\Sigma_b^{0})\simeq 6$ MeV and
$\Gamma (\Sigma_b^{*0})\simeq 10$ MeV predicted in~\cite{Liu:2012sj}, we estimate
that the branching fractions of $\mathcal{B}[\Sigma_b^{0}\to \Lambda_b^0\gamma]$
and $\mathcal{B}[\Sigma_b^{*0}\to \Lambda_b^0\gamma]$
can reach up to $2\%$ and $3\%$, respectively. It suggests that the missing neutral ground states $\Sigma_b^{0}$ and $\Sigma_b^{*0}$ may
be established in the $\Lambda_b^0\gamma$ channel.

\begin{table*}[htp]
\begin{center}
\caption{\label{Rad2}  Partial widths (keV) and branching fractions for the radiative decays of
the $1P$-wave states in the $\Sigma_c$ and $\Sigma_b$ families. }
\begin{tabular}{c|p{1.6cm}p{1.2cm}p{1.6cm}p{1.2cm}p{1.6cm}p{1.2cm}p{1.6cm}p{1.2cm}p{1.6cm}p{1.2cm}ccccccccccc|}\hline\hline
%                      &       &&         &     &  \underline{$1P$ $\Sigma_c$ states }      &   &&      &      \\
&\multicolumn{2}{c}{$\underline{~~|\Sigma_c~^2P_{\lambda} \frac{1}{2}^- \rangle(2713)~~}$}
&\multicolumn{2}{c}{$\underline{~~|\Sigma_c~^2P_{\lambda} \frac{3}{2}^- \rangle(2798)~~}$}
&\multicolumn{2}{c}{$\underline{~~|\Sigma_c~^4P_{\lambda} \frac{1}{2}^- \rangle(2799)~~}$}
&\multicolumn{2}{c}{$\underline{~~|\Sigma_c~^4P_{\lambda} \frac{3}{2}^- \rangle(2773)~~}$}
&\multicolumn{2}{c}{$\underline{~~|\Sigma_c~^4P_{\lambda} \frac{5}{2}^- \rangle(2789)~~}$} \\
                                 &~~~~~~$\Gamma_i$         &$\mathcal{B}_i(\%)$ &~~~~~~$\Gamma_i$         & $\mathcal{B}_i(\%)$&~~~~~~ $\Gamma_i$         & $\mathcal{B}_i(\%)$
                                 &~~~~~~$\Gamma_i$         & $\mathcal{B}_i(\%)$&~~~~~~ $\Gamma_i$         & $\mathcal{B}_i(\%)$\\ \hline
$\to\Sigma_c^{++}\gamma~$        &~~~283        &1.25        &~~~210      &0.58      &~~~8.54      &0.05     &~~~17.5      &0.07    &~~~13.6    &0.04   \\
$\to\Sigma_c^{+}\gamma~~ $       &~~~$1.60$     &$<0.01$     &~~~4.64     &0.01      &~~~0.92      &$<0.01$  &~~~1.86      &0.01    &~~~1.46     &$<0.01$   \\
$\to\Sigma_c^{0}\gamma~~ $       &~~~205        &0.91        &~~~245      &0.67      &~~~1.02      &$<0.01$  &~~~2.12      &0.01    &~~~1.64     &$<0.01$   \\
$\to\Lambda_c^{+}\gamma~~$       &~~~48.3       &0.21        &~~~87.3     &0.24      &~~~52.1      &0.30     &~~~105       &0.43    &~~~59.4     &0.18   \\
$\to\Sigma_c^{*++}\gamma $       &~~~3.04       &0.01        &~~~14.7     &0.04      &~~~387       &2.20     &~~~181       &0.73    &~~~168      &0.51   \\
$\to\Sigma_c^{*+}\gamma~$        &~~~0.31       &$<0.01$     &~~~1.55     &$<0.01$   &~~~1.75      &0.01     &~~~0.68      &$<0.01$ &~~~0.89     &$<0.01$   \\
$\to\Sigma_c^{*0}\gamma~$        &~~~0.39       &$<0.01$     &~~~1.82     &$<0.01$   &~~~289       &1.64     &~~~159       &0.65    &~~~160      &0.48   \\
\hline\hline
%                      &  &&              &     &  \underline{$1P$ $\Sigma_b$ states }      &   &&      &       \\
&\multicolumn{2}{c}{$\underline{~~|\Sigma_b~^2P_{\lambda} \frac{1}{2}^- \rangle(6101)~~}$}
&\multicolumn{2}{c}{$\underline{~~|\Sigma_b~^2P_{\lambda} \frac{3}{2}^- \rangle(6096)~~}$}
&\multicolumn{2}{c}{$\underline{~~|\Sigma_b~^4P_{\lambda} \frac{1}{2}^- \rangle(6095)~~}$}
&\multicolumn{2}{c}{$\underline{~~|\Sigma_b~^4P_{\lambda} \frac{3}{2}^- \rangle(6087)~~}$}
&\multicolumn{2}{c}{$\underline{~~|\Sigma_b~^4P_{\lambda} \frac{5}{2}^- \rangle(6084)~~}$} \\
&~~~~~~ $\Gamma_i$         &$\mathcal{B}_i(\%)$ &~~~~~~ $\Gamma_i$         & $\mathcal{B}_i(\%)$&~~~~~~ $\Gamma_i$         & $\mathcal{B}_i(\%)$
                                 &~~~~~~ $\Gamma_i$         & $\mathcal{B}_i(\%)$&~~~~~~ $\Gamma_i$         & $\mathcal{B}_i(\%)$\\ \hline
$ \rightarrow\Sigma_b^{+}\gamma$    &~~~1016   &4.49      &~~~483     &1.23      &~~~5.31    &0.04     &~~~13.1    &0.05      &~~~8.07  &0.02 \\
$ \rightarrow\Sigma_b^{0}\gamma$    &~~~74.9   &0.33      &~~~37.9    &0.10      &~~~0.32    &$<0.01$  &~~~0.80    &$<0.01$   &~~~0.49  &$<0.01$  \\
$ \rightarrow\Sigma_b^{-}\gamma$    &~~~212    &0.94      &~~~94.0    &0.24      &~~~1.37    &0.01     &~~~3.39    &0.01      &~~~2.08  &$<0.01$  \\
$ \rightarrow\Lambda_c^{0}\gamma$   &~~~133    &0.59      &~~~129     &0.33      &~~~63.6    &0.45     &~~~170     &0.65      &~~~83.3  &0.22  \\
$ \rightarrow\Sigma_b^{*+}\gamma$   &~~~16.9   &0.07      &~~~15.6    &0.04      &~~~867     &6.10     &~~~527     &2.00      &~~~426   &1.11  \\
$ \rightarrow\Sigma_b^{*0}\gamma$   &~~~1.03   &$<0.01$   &~~~0.95    &$<0.01$   &~~~63.6    &0.45     &~~~39.8    &0.15      &~~~32.6  &0.09  \\
$ \rightarrow\Sigma_b^{*-}\gamma$   &~~~4.36   &0.02      &~~~4.02    &0.01      &~~~182     &1.28     &~~~107     &0.41      &~~~85.3  &0.22  \\
\hline\hline
\end{tabular}
\end{center}
\end{table*}

\subsubsection{$1P$ states}\label{PP}

In the $\Sigma_c$ and $\Sigma_b$ families, there are five $\lambda$-mode $1P$-wave excitations: $|^2P_{\lambda} \frac{1}{2}^- \rangle$,
$|^4P_{\lambda} \frac{1}{2}^- \rangle$, $|^2P_{\lambda} \frac{3}{2}^- \rangle$,
$|^4P_{\lambda} \frac{3}{2}^- \rangle$ and $|^4P_{\lambda} \frac{5}{2}^- \rangle$ within the quark model.
However, so far there are no $1P$-wave states indisputably established.
The masses of the $\lambda$-mode $1P$-wave $\Sigma_c$ and $\Sigma_b$ excitations
are predicted to be $\sim 2.8$ and $\sim 6.1$ GeV, respectively, within various quark models (see Table~\ref{sp1}).
We study the two body OZI-allowed strong decays of the $\lambda$-mode $1P$-wave $\Sigma_c$
and $\Sigma_b$ states in their possible mass ranges. Our results are
shown in Fig.~\ref{1pscb}. To be more specific, we assume that the mass ranges of the $\lambda$-mode $1P$ states can be reasonably constrained although their ordering and detailed spectrum could be model dependent. Within a restricted mass region, the partial decay widths would be more sensitive to the detailed dynamics. Thus, by taking the masses of the $1P$-wave states
compatible with those from RQM~\cite{Ebert:2011kk}, the calculated results listed
in Table~\ref{sigemac} are to be compared with other model calculations.

From Fig.~\ref{1pscb} and Table~\ref{sigemac}, one can see that
in the $\Sigma_c$ and $\Sigma_b$ families,
both the $|^2P_{\lambda} \frac{1}{2}^- \rangle$ and $|^4P_{\lambda} \frac{1}{2}^- \rangle$
excitations may be narrow states with a width of $\sim 10-20$ MeV. The decays of
$|\Sigma_{c(b)}~^2P_{\lambda} \frac{1}{2}^- \rangle$ are governed by the $\Sigma_{c(b)} \pi$ channel,
while $|\Sigma_{c(b)}~^4P_{\lambda} \frac{1}{2}^- \rangle$ may mainly decay into
$\Lambda_{c(b)} \pi$ and $\Sigma_{c(b)} \pi$ channels.

Furthermore, one can see that the spin-3/2 states with $J^P=3/2^-$,
$|\Sigma_c~^4P_{\lambda} \frac{3}{2}^- \rangle$
and $|\Sigma_b~^4P_{\lambda} \frac{3}{2}^- \rangle$, mainly decay
into the $\Sigma_c^*\pi$ and $\Sigma_b^*\pi$ channels, respectively.
They might be narrow states with a width of $\Gamma\simeq 30$ MeV.
However, for the other $P$-wave $\Sigma_c$ ($\Sigma_b$) states,
the decay rates into the $\Sigma_c^*\pi$ ($\Sigma_b^*\pi$)
channel is very small. Thus, a narrow resonance around $\sim 2.8$ GeV and dominantly decaying into
$\Sigma_c^*\pi$ should be a good signal for the $|\Sigma_c~^4P_{\lambda} \frac{3}{2}^- \rangle$ state.
Similar criterion can be applied to the $|\Sigma_b~^4P_{\lambda} \frac{3}{2}^- \rangle$ state.

It shows that both $|^2P_{\lambda} \frac{3}{2}^- \rangle$ and
$|^4P_{\lambda} \frac{5}{2}^- \rangle$ excitations have similar
widths of $\sim 40$ MeV in the $\Sigma_c$ and $\Sigma_b$ spectra,
and their main strong decay modes are similar to each other. Thus,
to distinguish $|^2P_{\lambda} \frac{3}{2}^- \rangle$ from
$|^4P_{\lambda} \frac{5}{2}^- \rangle$, additional information apart from the strong decay properties is needed in order to determine their quantum numbers, such as the angle distributions or radiative decay properties.

$\Sigma_c(2800)$ might be a good candidate of
the $\lambda$-mode $1P$-wave excitations. This state was firstly observed
in the $\Lambda_c^+ \pi$ channel by Belle~\cite{Mizuk:2004yu};
its measured width is $\sim 70$ MeV with a large uncertainty~\cite{Olive:2016xmw}.
Some predictions about its nature can be found in the literature~\cite{Cheng:2006dk,Chen:2007xf,Ebert:2007nw,
Gerasyuta:2007un,Garcilazo:2007eh}; however, there are strong model dependencies.
For example, its spin-parity numbers were suggested to be $J^P=3/2^-$ in HHChPT~\cite{Cheng:2006dk};
$J^P=3/2^-$ or $J^P=5/2^-$ in the $^3P_0$ model~\cite{Chen:2007xf},
$J^P=5/2^-$ in RQM~\cite{Gerasyuta:2007un}, and $J^P=1/2^-$ or $3/2^-$ in the Faddeev
studies~\cite{Garcilazo:2007eh}. In this work, we revised our predictions of $\Sigma_c(2800)$.
It shows that $\Sigma_c(2800)$ may favor the $|\Sigma_c~ ^2P_{\lambda} \frac{3}{2}^- \rangle$ or the $J^P=5/2^-$ state
$|\Sigma_c~ ^4P_{\lambda} \frac{5}{2}^- \rangle$ assignments (see Table~\ref{sigemac} and Fig.~\ref{1pscb}).
Both of them have comparable decay widths of $\Gamma\sim 40$ MeV,
and their decays are dominated by the $\Lambda_c\pi$ mode. These features
are consistent with the observations of $\Sigma_c(2800)$. It should be pointed out that $\Sigma_c(2800)$ may
be caused by two largely overlapping states $|\Sigma_c~ ^2P_{\lambda} \frac{3}{2}^- \rangle$
and $|\Sigma_c~ ^4P_{\lambda} \frac{5}{2}^- \rangle$ at $\sim 2.8$ GeV.
We still need more observables to clarify its nature.

The radiative decays of the $1P$ $\Sigma_c$ and $\Sigma_b$ states can provide more information about their internal structures.
Our calculations of their radiative decays into $1S$ states are listed in Table~\ref{Rad2}.
Some of the $1P$ states indeed appear to have large radiative decay rates into the
$1S$-wave states. For example, the decay of $|\Sigma_c~^2P_{\lambda} \frac{3}{2}^- \rangle \to\Sigma_c^{++,0}\gamma$
has a branching fraction up to $\mathcal{O}(10^{-2})$. Also, the branching ratios of $|\Sigma_c~^4P_{\lambda} \frac{3}{2}^- \rangle$
and $|\Sigma_c~^4P_{\lambda} \frac{5}{2}^- \rangle\to \Sigma_c^{*++,0}\gamma$ can reach up to
$\mathcal{O}(10^{-2})$. These are potential processes that can be accessed in experiment. It shows that the branching fraction for the singly charged $1P$-wave $\Sigma_c^+\Lambda_c^+\gamma$ is at $\mathcal{O}(10^{-3})$ which may still be accessible in an experiment.

Similarly, in the $1P$-wave $\Sigma_b$ sector the branching fractions for $|\Sigma_b^+~^2P_{\lambda} \frac{3}{2}^- \rangle\to \Sigma_b^{+}\gamma$, and $|\Sigma_b^+~^4P_{\lambda} \frac{3}{2}^- \rangle$
and $|\Sigma_b^+~^4P_{\lambda} \frac{5}{2}^- \rangle\to \Sigma_b^{*+}\gamma$ reach up to $\mathcal{O}(10^{-2})$, which suggests that they can be searched in these radiative decays in future experiment.

\begin{figure}[h]
\begin{center}
\centering  \epsfxsize=7.2 cm \epsfbox{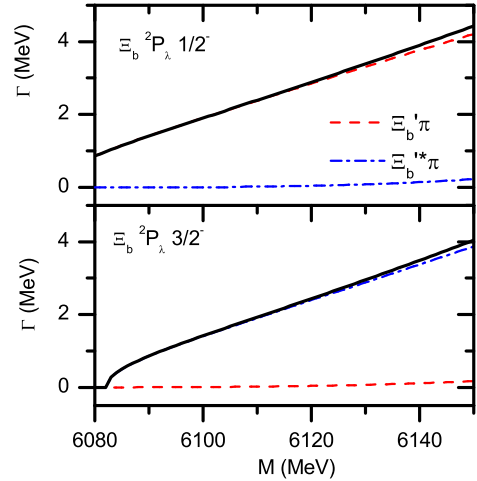} %\epsfxsize=5.0 cm \epsfbox{1DXB.eps}
\vspace{-0.3 cm} \caption{Partial and total strong decay widths of the $1P$-wave $\Xi_b$ states
as functions of their mass. The solid curves stand for the total widths.  } \label{PXib}
\end{center}
\end{figure}

\subsection{The $\Xi_c$  and $\Xi_b$ states}

In the $\Xi_c$ and $\Xi_b$ families,
there are two $\lambda$-mode $1P$ excitations $|^2P_\lambda \frac{1}{2}^-\rangle$
and $|^4P_\lambda \frac{3}{2}^-\rangle$ according to the quark model classification (see Table~\ref{sp1}).
In our previous analysis, we find that the well-established states $\Xi_c(2790)$ and $\Xi_c(2815)$
could be assigned to the two low-lying $1P$ states with
$J^P=1/2^-$ and $3/2^-$, respectively~\cite{Liu:2012sj}.
Very recently, the Belle Collaboration reported the accurate measurements
of the widths of $\Xi_c(2790)^{0,+}$ and $\Xi_c(2815)^{0,+}$~\cite{Yelton:2016fqw},
which allows us to revisit the assignments of $\Xi_c(2790)$ and $\Xi_c(2815)$.
Our calculation results are listed in Table~\ref{Xicb}. It shows that by assigning the $\Xi_c(2815)$
as the $\lambda$-mode $1P$ excited state with $J^P=3/2^-$, the obtained
width, $\Gamma=2.1$ MeV, is in good agreement with the data, $2.5$ MeV, from
Belle~\cite{Yelton:2016fqw}. However, if assigning the $\Xi_c(2790)$
as the $\lambda$-mode $1P$ excited state with $J^P=1/2^-$, we find that
the theoretical width $\Gamma=3.6$ MeV is about a factor of $3$ smaller
than the measured value $\sim 10$ MeV from Belle~\cite{Yelton:2016fqw}.
It should be mentioned that the predicted widths of $\Xi_c(2790)$
within the $^3P_0$ model~\cite{Chen:2007xf} and HHChPT~\cite{Cheng:2006dk}
are also close to the measured value. In contrast, there exist significant discrepancies
between the theoretical calculations and measured value for  $\Xi_c(2815)$. In Ref.~\cite{Chen:2015kpa},
the authors proposed that the $\Xi_c(2790)$ and $\Xi_c(2815)$ may be assigned to
the $\rho$-mode excitations within their QCD sum rule analysis. However, in such a case as the
$\rho$-mode excitations, their strong decays should be dominated by the
$\Lambda_cK$ and $\Xi_c\pi$ channels and the $\Xi'_c\pi$ channel
would be forbidden~\cite{Liu:2012sj}.

To know more about the properties of $\Xi_c(2790)$ and $\Xi_c(2815)$,
we further study their radiative transitions and the results are listed in Table~\ref{rad2}.
Considering $\Xi_c(2790)$ and $\Xi_c(2815)$ as the $\lambda$-mode $1P$ excitations
$|\Xi_c~^2P_\lambda \frac{1}{2}^-\rangle$ and $|\Xi_c~^4P_\lambda \frac{3}{2}^-\rangle$,
we find that both $\Xi_c(2790)^0$ and $\Xi_c(2815)^0$ have a large
decay rate into $\Xi_c^0\gamma$.
The partial widths are predicted to be
\begin{eqnarray}\label{1xc}
\Gamma[\Xi_c(2790)^0\to \Xi_c^0\gamma]&\simeq & 263 \ \mathrm{keV},\\
\Gamma[\Xi_c(2815)^0\to \Xi_c^0\gamma]&\simeq & 292 \ \mathrm{keV}.
\end{eqnarray}
The partial width of $\Gamma[\Xi_c(2815)^0\to \Xi_c^0\gamma]$ predicted by us
is consistent with that of the RQM result~\cite{Ivanov:1999bk}.
Combining the measured widths of $\Xi_c(2790)^0$ and $\Xi_c(2815)^0$,
we obtain fairly large branching fractions
\begin{eqnarray}
\mathcal{B}[\Xi_c(2790)^0\to \Xi_c^0\gamma]&\simeq & 3\%,\\
\mathcal{B}[\Xi_c(2815)^0\to \Xi_c^0\gamma]&\simeq & 12\%,
\end{eqnarray}
which indicates that $\Xi_c(2790)^0\to \Xi_c^0\gamma$ and
$\Xi_c(2815)^0\to \Xi_c^0\gamma$ should be ideal for future experimental searches.
For the charged state $\Xi_c(2815)^+$, we predict a small decay rate into
$\Xi_c^+\gamma$, i.e., $\Gamma[\Xi_c(2815)^+\to \Xi_c^+\gamma]\simeq 3$ keV,
which is about two orders of magnitude smaller than that from
Ref.~\cite{Ivanov:1999bk}. The radiative decay mechanism for
$\Xi_c(2815)^+\to \Xi_c^+\gamma$ is the same as that for
$\Lambda_c(2625)\to \Lambda_c^+\gamma$, which has been discussed in
Sec.~\ref{LLL}.

Furthermore, considering $\Xi_c(2790)$ and $\Xi_c(2815)$ as the $P$-wave $\rho$-mode
excitations, we also study their radiative decays
within the constituent quark model. The results are listed
in Table~\ref{rad2} as well [denoted by ours (B), (C)].
Again, we see the difference between the $\rho$-mode and $\lambda$-mode excitations in the radiative decays.
Future measurements of $\Xi_c(2790)^0\to \Xi_c^0\gamma$ and
$\Xi_c(2815)^0\to \Xi_c^0\gamma$ will allow us to test various model predictions and better understand
the internal structures of these two states.

In the $\Xi_b$ baryon sector, no $1P$ states have been observed in an experiment.
The typical masses of the $\lambda$-mode
$1P$-wave $\Xi_b$ excitations are $\sim 6.1$ GeV
from various quark model predictions (see Table~\ref{sp1}). Our results for the strong decays of the $\lambda$-mode $1P$ excitations $|\Xi_b~^2P_\lambda \frac{1}{2}^-\rangle$
and $|\Xi_b~^2P_\lambda \frac{3}{2}^-\rangle$ are shown in
Fig.~\ref{PXib}. The strong decays of $|\Xi_b~^2P_\lambda \frac{1}{2}^-\rangle$ and
$|\Xi_b~^2P_\lambda \frac{3}{2}^-\rangle$ appear to be dominated by the $\Xi_b'\pi$ and $\Xi_b'^*\pi$ channels, respectively.
If we take a typical theoretical mass $\sim6.12$ GeV (for instance, from RQM~\cite{Ebert:2011kk}) for these two
$1P$ excited $\Xi_b$ states, we find that they should have a very narrow width of $\Gamma\simeq 3$ MeV
(see Table~\ref{Xicb}).

The radiative transitions of $|\Xi_b~^2P_\lambda \frac{1}{2}^-\rangle$
and $|\Xi_b~^2P_\lambda \frac{3}{2}^-\rangle\to \Xi_b^{0,-} \gamma$ may play a crucial role in their decays because of
their narrow widths. Sizeable partial decay widths for these two states are obtained in the calculation and our results are listed in Table~\ref{rad2}. Combining the predicted total widths, we find that the branching fractions
for these two $1P$ state radiative decays into
$\Xi_b^{0,-} \gamma$ can reach up to $\sim 2\%$ if they are $\lambda$-mode excitations indeed.
This makes the $\Xi_b^{0,-} \gamma$ channel
an ideal place for the search of these two states in future experiments.

\begin{table*}[ht]
\begin{center}
\caption{ \label{Xicb} Partial widths (MeV) and branching fractions for the strong decays of the $1P$-wave states in the $\Xi_c$ and $\Xi_b$ families.}
%\footnotesize
\begin{tabular}{p{1.5cm}p{1.5cm}p{2.0cm}p{1.5cm}p{1.5cm}p{1.5cm}p{2.0cm}p{1.5cm}p{1.5cm}p{1.5cm}cccccc}
\hline\hline
$|^{2S+1}L_\lambda~J^P\rangle$  &State      &Channel         &$\Gamma_{i}$ (MeV)    &$\mathcal{B}_i$     &State      &Channel         &$\Gamma_{i}$ (MeV)     &$\mathcal{B}_i$   \\
\hline
$|^2P_{\lambda} \frac{1}{2}^- \rangle$   &$\Xi_c(2790)$     &$\Xi'_c\pi$      &3.61   &100\%    &$\Xi_b(6120)$  &$\Xi'_b\pi$   &2.84 &98.61\%\\
                    &                                       &$\Xi'^*_c\pi$           &$3.9\times10^{-4}$ &$\simeq0.0$\%  &                 &$\Xi'^*_b\pi$   &0.04         &1.39\% \\
                    &                                       &total                  &3.61               &               &                  &total         &2.88               & \\
\hline
$|^2P_{\lambda} \frac{3}{2}^- \rangle$   &$\Xi_c(2815)$     &$\Xi'_c\pi$            &0.31               &14.69\%        &$\Xi_b(6130)$     &$\Xi'_b\pi$    &0.07          &2.37\%\\
                        &                                   &$\Xi^*_c\pi$           &1.80               &85.31\%         &                 &$\Xi'^*_b\pi$   &2.88         &97.63\%\\
                    &                                       &total                  & 2.11              &                &                  &total         &2.95               &        \\
\hline\hline
\end{tabular}
\end{center}
\end{table*}

\begin{table*}[htb]
\begin{center}
\caption{ \label{Xipcb} Partial widths (MeV) and branching fractions for the strong decays of the $1P$-wave states in the $\Xi_c'$ and $\Xi_b'$ families.}
%\footnotesize
\begin{tabular}{p{1.5cm}p{1.5cm}p{2.0cm}p{1.5cm}p{1.5cm}p{1.5cm}p{2.0cm}p{1.5cm}p{1.5cm}p{1.5cm}cccccccccccc}
\hline\hline
$|^{2S+1}L_\lambda~J^P\rangle$  & State      &Channel         &$\Gamma_{i}$ (MeV)    &$\mathcal{B}_i$   & State      &Channel         &$\Gamma_{i}$ (MeV)    &$\mathcal{B}_i$   \\
\hline
%$|^2S\frac{3}{2}^+ \rangle$              &$\Xi'_c(5945)$       &$\Xi_c^0\pi^+$      &1.58        &61.96\%        &                    \\
%                        &                                      &$\Xi_c^{+}\pi^0$    &0.81        &31.76\%        &                \\
%                        &                                      &total               &2.55        &              &                 \\
%\hline
$|^2P_{\lambda} \frac{1}{2}^- \rangle$   &$\Xi'_c(2936)$       &$\Lambda_cK$        &7.11        &32.81\%      &$\Xi'_b(6233)$   &$\Lambda_bK$        &12.11        &44.77\%                \\
                        &                                      &$\Xi_c\pi$          &3.90        &18.00\%      &                 &$\Xi_b\pi$          &4.77        &17.63\%   \\
                        &                                &$\Xi^\prime_c(2580)\pi$   &10.08       &46.52\%      &           &$\Xi^\prime_b\pi$         &9.23       &34.12\% \\
                        &                                &$\Xi^\prime_c(2645)\pi$   &0.58        &2.68\%       &           &$\Xi^\prime_b(5945)\pi$   &0.94        &3.48\%\\
                        &                                      &total               &21.67       &             &           &total                     &27.05       &            \\
\hline
$|^2P_{\lambda} \frac{3}{2}^- \rangle$   &$\Xi'_c(2935)$        &$\Lambda_cK$       &3.73        &17.86\%      &$\Xi'_b(6234)$   &$\Lambda_bK$        &4.14        &17.14\%              \\
                        &                                       &$\Xi_c\pi$         &10.85       &51.94\%      &                 &$\Xi_b\pi$          &14.91        &61.74\%             \\
                        &                                       &$\Xi'_c(2580)\pi$  &3.89        &18.62\%      &           &$\Xi^\prime_b\pi$         &2.37       &9.81\%        \\
                        &                                       &$\Xi'_c(2645)\pi$  &2.42        &11.58\%      &           &$\Xi^\prime_b(5945)\pi$   &2.73        &11.30\%       \\
                        &                                       &total              &20.89       &              &                  &total              &24.15       & \\
\hline
$|^4P_{\lambda} \frac{1}{2}^- \rangle$   &$\Xi'_c(2854)$        &$\Lambda_cK$       &18.56       &50.09\%       &$\Xi'_b(6227)$   &$\Lambda_bK$        &17.28        &53.60\%         \\
                        &                                       &$\Xi_c\pi$         &15.02       &40.54\%       &                 &$\Xi_b\pi$          &10.01        &31.05\%         \\
                        &                                       &$\Xi'_c(2580)\pi$  &3.44        &9.28\%        &           &$\Xi^\prime_b\pi$         &4.54       &14.08\%\\
                        &                                       &$\Xi'_c(2645)\pi$  &0.03        &0.07              &           &$\Xi^\prime_b(5945)\pi$   &0.41        &1.27\%       \\
                        &                                       &total              &37.05       &              &                   &total              &32.24       &  \\
\hline
$|^4P_{\lambda} \frac{3}{2}^- \rangle$   &$\Xi'_c(2912)$        &$\Lambda_cK$       &0.50        &4.06\%        &$\Xi'_b(6224)$   &$\Lambda_bK$        &0.98        &6.19\%         \\
                        &                                       &$\Xi_c\pi$         &1.70        &13.79\%       &                 &$\Xi_b\pi$          &2.67        &16.87\% \\
                        &                                       &$\Xi'_c(2580)\pi$  &0.13        &1.05\%        &           &$\Xi^\prime_b\pi$         &0.10       &0.63\%\\
                        &                                       &$\Xi'_c(2645)\pi$  &10.00       &81.10\%       &           &$\Xi^\prime_b(5945)\pi$   &12.08        &76.31\%      \\
                        &                                       &total              &12.33       &              &                   &total              &15.83       &  \\
\hline
$|^4P_{\lambda} \frac{5}{2}^- \rangle$   &$\Xi'_c(2929)$        &$\Lambda_cK$       &4.06        &20.10\%       &$\Xi'_b(6226)$   &$\Lambda_bK$        &4.20        &17.22\%         \\
                        &                                       &$\Xi_c\pi$         &12.24       &60.59\%       &                 &$\Xi_b\pi$          &16.37        &67.12\%\\
                        &                                       &$\Xi'_c(2580)\pi$  &1.06        &5.25\%        &           &$\Xi^\prime_b\pi$         &0.60       &2.46\%\\\
                        &                                       &$\Xi'_c(2645)\pi$  &2.84        &14.06\%       &           &$\Xi^\prime_b(5945)\pi$   &3.22        &13.20\%\\
                        &                                                 &total    &20.2       &              &                   &total             &24.39       &  \\
\hline\hline
\end{tabular}
\end{center}
\end{table*}

\subsection{The $\Xi'_c$  and $\Xi'_b$ states}

\subsubsection{$1S$ states}

In the $\Xi'_c$ family, the ground $1S$-wave states
with $J^P=1/2^+$ and $J^P=3/2^+$ have been established. They
correspond to $\Xi'_c$ and $\Xi'_c(2645)$ (denoted by $\Xi'^{*}_c$)
listed in the PDG book~\cite{Olive:2016xmw}, respectively. In~\cite{Liu:2012sj},
we studied the strong decays of $\Xi'_c(2645)$ in the ChQM. The theoretical width $\Gamma\simeq 2.4$ MeV is
in good agreement with the recent measured value from Belle~\cite{Yelton:2016fqw}.
To better understand the properties of $\Xi'_c$ and $\Xi'_c(2645)$
and test our model, we further study their radiative
decays, and our results are listed in Table~\ref{Rad}. It shows that the charged $\Xi'_c$ and $\Xi'_c(2645)$ have
large radiative decay rates into $\Xi_c^+\gamma$ and the corresponding partial decay widths read
\begin{eqnarray}\label{1ss11}
\Gamma[\Xi'^{+}_c\to \Xi_c^+\gamma]&\simeq & 42 \ \mathrm{keV}, \\
\Gamma[\Xi'^{*+}_c\to \Xi_c^+\gamma]&\simeq & 139 \ \mathrm{keV},
\end{eqnarray}
which are consistent with other model calculations (see Table~\ref{Rad}). The
$\Xi'^{+}_c\to \Xi_c^+\gamma$ process has been seen by CLEO~\cite{Jessop:1998wt}. While the branching ratios of
$\mathcal{B}[\Xi'_c(2645)^+\to \Xi_c^+\gamma]$ can reach up to $4\%$, it can be searched in a future experiment.
It should be mentioned that we may also observe the M1 transition
$\Xi'_c(2645)^0\to \Xi_c'^0\gamma$ in a future experiment due to
its sizeable partial width and branching ratio
\begin{eqnarray}\label{1ss11}
\Gamma[\Xi'_c(2645)^0\to \Xi_c'^0\gamma]&\simeq & 3.0 \ \mathrm{keV}, \\
\mathcal{B}[\Xi'_c(2645)^0\to \Xi_c'^0\gamma]& \simeq & 1.2\times 10^{-3}.
\end{eqnarray}

In the $\Xi'_b$ family, except the neutral ground state $\Xi'^{0}_b$,
all the $1S$-wave ground states with $J^P=1/2^+$ and $J^P=3/2^+$ have been observed at LHC
during the past five years~\cite{Chatrchyan:2012ni,Aaij:2014yka}.
In 2012, the CMS Collaboration firstly observed
a neutral state, $\Xi_b(5945)^0$,  in the $\Xi_b^-\pi^+$ channel, which was naturally explained as
the $J^P=3/2^+$ ground state $\Xi'^{*0}_b$~\cite{Chatrchyan:2012ni}.  Recently,
the LHCb Collaboration observed two new charged states $\Xi'_b(5935)^-$
and $\Xi_b'^*(5955)^-$ in the $\Xi^0_b\pi^-$ mass spectrum~\cite{Aaij:2014yka}.
They were proposed to be the $J^P=1/2^+$ and $J^P=3/2^+$ ground states
$\Xi'^{-}_b$ and $\Xi'^{*-}_b$, respectively. The decays of the $J^P=3/2^+$
states $\Xi'^{*0}_b$ and $\Xi'^{*-}_b$ are dominated by $\Xi_b\pi$.
In the ChQM, we obtain
\begin{eqnarray}\label{ss32}
\Gamma[\Xi'^{*0}_b\to \Xi_b\pi]&\simeq & 0.73 \ \mathrm{MeV}, \\
\Gamma[\Xi'^{*-}_b\to \Xi_b\pi]& \simeq & 1.23\ \mathrm{MeV},
\end{eqnarray}
which are in good agreement with the recently measured values
$\Gamma(\Xi'^{*0}_b)\simeq 0.90\pm 0.24$ MeV~\cite{Aaij:2016jnn} and
$\Gamma(\Xi'^{*-}_b)\simeq 1.65\pm 0.41$ MeV~\cite{Aaij:2014yka}
from the LHCb Collaboration. The $J^P=1/2^+$ state $\Xi'_b(5935)^-$
is very close to the threshold of $\Xi_b\pi$. Thus, its decays into $\Xi_b\pi$
are strongly suppressed by the phase space. In the ChQM,
we obtain a very small width
\begin{eqnarray}\label{ss12}
\Gamma[\Xi'^{-}_b\to \Xi_b\pi]&\simeq & 78 \ \mathrm{keV},
\end{eqnarray}
which is close to the upper limit of $\Gamma(\Xi'^{-}_b)=80$ keV
measured by LHCb~\cite{Aaij:2014yka}.

The mass of $\Xi'^{0}_b$ is estimated to be about 5929 MeV with the measured
isospin splitting $m(\Xi^{-}_b)-m(\Xi^{0}_b)\simeq 6$ MeV
~\cite{Aaij:2014lxa}. The mass of $\Xi'^{0}_b$ is below the
threshold of $\Xi_b\pi$; thus, $\Xi'^{0}_b$ should decay
electromagnetically via the process $\Xi'^{0}_b\to \Xi_b^0\gamma$.
Its partial width is estimated to be
\begin{eqnarray}\label{radxb}
\Gamma[\Xi'^{0}_b\to \Xi_b^0\gamma]&\simeq & 85 \ \mathrm{keV}.
\end{eqnarray}

Furthermore, we estimate the radiative decays of $\Xi'^{*0,-}$ and the results
are listed in Table~\ref{Rad}. It shows that the
$\Xi'^{*0}$ and $\Xi'^{*-}$ have relatively large decay rates
into $\Xi_b^0\gamma$ and $\Xi_b'^-\gamma$, respectively.
It should be mentioned that the M1 transition
$\Xi'^{*0}\to \Xi_b'^0\gamma$ has sizeable partial width and branching fraction
\begin{eqnarray}\label{1ss11}
\Gamma[\Xi'^{*0}_b\to \Xi_b'^0\gamma]&\simeq & 5 \ \mathrm{keV}, \\
\mathcal{B}[\Xi'^{*0}_b\to \Xi_b'^0\gamma]& \simeq & 6\%.
\end{eqnarray}
Thus, it can be searched in future experiments.

%\begin{figure}[h]
%\begin{center}
%\centering  \epsfxsize=8.6 cm \epsfbox{1PXmix.eps}
%\vspace{-0.7 cm} \caption{Partial and total strong decay widths of the $J^P=1/2^-$ mixed states in the $\Xi_c'$ and $\Xi_b'$
%families as functions of their mass. The solid curves stand for the total widths.} \label{Xicbmi}
%\end{center}
%\end{figure}

%\begin{figure}[h]
%\begin{center}
%\centering  \epsfxsize=6.3 cm \epsfbox{1PMXc.eps}
%\vspace{-0.7 cm} \caption{Partial and total strong decay widths of the $J^P=1/2^-$ mixed states of $\Xi_c'$
%as functions of the mixing angle $\phi$. The masses of the mixed states are fixed with $M=2930$ MeV.
%The solid curves stand for the total widths.} \label{Xicbcc}
%\end{center}
%\end{figure}

\begin{figure}[htbp]
\begin{center}
\centering \epsfxsize=7.6 cm \epsfbox{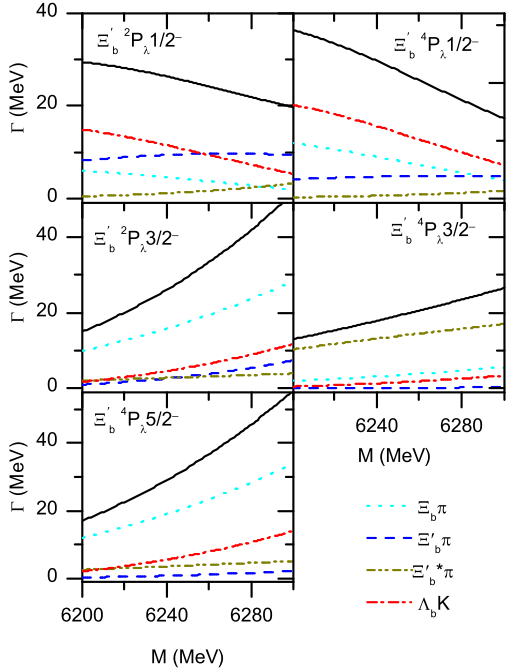}
\vspace{-0.5 cm} \caption{Partial and total strong decay widths of the $1P$ states in the $\Xi_b'$
family as functions of their mass. The solid curves stand for the total widths. } \label{Xippd}
\end{center}
\end{figure}

\subsubsection{$1P$ states}

In the $\Xi'_c$ family, the typical masses of the $\lambda$-mode $1P$-wave states $|\Xi_c'~^2P_\lambda\frac{1}{2}^-\rangle$,
$|\Xi_c'~^2P_\lambda\frac{3}{2}^-\rangle$, $|\Xi_c'~^4P_\lambda\frac{1}{2}^-\rangle$,
$|\Xi_c'~^4P_\lambda\frac{3}{2}^-\rangle$ and $|\Xi_c'~^4P_\lambda\frac{5}{2}^-\rangle$
are predicted to be around $\sim 2.93$ GeV by various quark model predictions (see Table~\ref{sp1}).
So far, no $P$-wave states have been established in an experiment.
It should be mentioned that in 2007 the Babar Collaboration observed some signals of
a neutral charmed-strange resonance, i.e. $\Xi_c(2930)^0$, in the $\Lambda_c^+K^-$ final state with a width of
$\Gamma\simeq 36\pm 18$ MeV~\cite{Aubert:2007eb}, which might be a good candidate for the $\lambda$-mode
$1P$-wave $\Xi_c'$ states. In~\cite{Liu:2012sj}, we studied the two-body
OZI-allowed strong decays of the $\lambda$-mode
$1P$-wave $\Xi_c'$ states in their possible mass ranges.
The observed decay mode $\Lambda_c^+K^-$ and decay
width $\Gamma\simeq 36\pm 18$ MeV indicate that $\Xi_c(2930)^0$ may favor the
$1P$-wave $\Xi_c'$ state with $J^P=1/2^-$~\cite{Liu:2012sj}.
The $\Xi_c(2930)$ resonance is expected to be confirmed by other experiments.

From Table \ref{Xipcb} it shows that $|\Xi_c'~^2P_\lambda\frac{3}{2}^-\rangle$
and $|\Xi_c'~^4P_\lambda\frac{5}{2}^-\rangle$
have compatible widths of $\Gamma\simeq 21$ MeV, and their decays are dominated by
the $\Xi_c\pi$ mode. Thus, to distinguish these two states additional information such as angular distributions may be needed in order to determine their quantum numbers. The other $J^P=3/2^-$ state $|\Xi_c'~^4P_\lambda\frac{3}{2}^-\rangle$ has
a narrow width of $\Gamma \simeq 12$ MeV, and mainly decays into the $\Xi_c'^{*}\pi$ channel. It makes
the $\Xi_c'^{*}\pi$ channel ideal for its search in experiments.

In the $\Xi'_b$ family, the typical masses of
the $\lambda$-mode $1P$-wave states $|\Xi'_b~^2P_\lambda\frac{1}{2}^-\rangle$,
$|\Xi'_b~^2P_\lambda\frac{3}{2}^-\rangle$, $|\Xi'_b~^4P_\lambda\frac{1}{2}^-\rangle$,
$|\Xi'_b~^4P_\lambda\frac{3}{2}^-\rangle$, and $|\Xi'_b~^4P_\lambda\frac{5}{2}^-\rangle$
are predicted to be around $\sim 6.23$ GeV (see Table~\ref{sp1}).
So far, there are no experimental data available for these states. In Fig.~\ref{Xippd} our calculations of their strong decay widths as function of
mass are plotted in Fig.~\ref{Xippd}. To be more specific, since the masses from different models do not drastically scatter to a broad range, we adopt the masses, e.g. from the relativistic quark-diquark picture~\cite{Ebert:2011kk}, to extract the partial decay widths. The results are listed
in Table~\ref{Xipcb}. It shows that the $\lambda$-mode $1P$-wave
$\Xi'_b$ states have a narrow width of $20\sim 30$ MeV.

The decays of $|\Xi_b'~^4P_\lambda\frac{3}{2}^-\rangle$ are governed by
the $\Xi_b'^*\pi$ channel, which makes the $\Xi_b'\pi$ channel interesting for future studies.
Both $|\Xi_b'~^2P_\lambda\frac{3}{2}^-\rangle$ and $|\Xi_b'~^4P_\lambda\frac{5}{2}^-\rangle$
have a similar decay width and their decays are dominated by the
same channel $\Xi_b\pi$. Therefore, in order to distinguish them,
more information, such as the angular distributions or radiative decay properties, should be
collected in experiments.

\begin{table*}[htp]
\begin{center}
\caption{\label{raXipp}  Partial widths (keV) and branching fractions for the radiative decays of the $\Xi^\prime_c$ and  $\Xi^\prime_b$  baryons.}
\begin{tabular}{c|p{1.6cm}p{1.2cm}p{1.6cm}p{1.2cm}p{1.6cm}p{1.2cm}p{1.6cm}p{1.2cm}p{1.6cm}p{1.2cm}ccccccccccccccccccccccccc|}\hline\hline
&\multicolumn{2}{c}{$\underline{~~|\Xi^\prime_c~^2P_{\lambda} \frac{1}{2}^- \rangle(2936)~~}$}
&\multicolumn{2}{c}{$\underline{~~|\Xi^\prime_c~^2P_{\lambda} \frac{3}{2}^- \rangle(2935)~~}$}
&\multicolumn{2}{c}{$\underline{~~|\Xi^\prime_c~^4P_{\lambda} \frac{1}{2}^- \rangle(2854)~~}$}
&\multicolumn{2}{c}{$\underline{~~|\Xi^\prime_c~^4P_{\lambda} \frac{3}{2}^- \rangle(2912)~~}$}
&\multicolumn{2}{c}{$\underline{~~|\Xi^\prime_c~^4P_{\lambda} \frac{5}{2}^- \rangle(2929)~~}$} \\
                                 &~~~~~~ $\Gamma_i$         &$\mathcal{B}_i(\%)$ & $~~~~~~\Gamma_i$         & $\mathcal{B}_i(\%)$& $~~~~~~\Gamma_i$ & $\mathcal{B}_i(\%)$
                                 &~~~~~~ $\Gamma_i$         & $\mathcal{B}_i(\%)$& $~~~~~~\Gamma_i$         & $\mathcal{B}_i(\%)$\\ \hline
$  \to         \Xi_c^{+}\gamma$         &~~~46.4   &0.21        &~~~46.1   &0.22      &~~~14.5  &0.04       &~~~54.6   &0.44     &~~~32.0   &0.16   \\
$  \to         \Xi_c^{0}\gamma$         &~~~0.0    &0.0         &~~~0.0    &0.0       &~~~0.00  &0.00       &~~~0.00   &0.0      &~~~0.00   &0.00   \\
$  \to   \Xi_c^{\prime+}\gamma$         &~~~0.03   &$<0.01$     &~~~12.1   &0.06      &~~~0.33  &$<0.01$    &~~~2.06   &0.02     &~~~1.63   &$<0.01$   \\
$  \to   \Xi_c^{\prime0}\gamma$         &~~~472    &2.18        &~~~302    &1.45      &~~~0.20  &$<0.01$    &~~~1.21   &$<0.01$  &~~~0.93   &$<0.01$   \\
$  \to   \Xi_c^{*\prime+}\gamma$        &~~~1.61   &$<0.01$     &~~~1.59   &$<0.01$   &~~~0.16  &$<0.01$    &~~~1.64   &0.01     &~~~2.35   &0.01   \\
$  \to   \Xi_c^{*\prime0}\gamma$        &~~~1.00   &$<0.01$     &~~~1.05   &$<0.01$   &~~~125   &0.34       &~~~187    &1.52     &~~~192    &0.95    \\
\hline\hline
&\multicolumn{2}{c}{$\underline{~~|\Xi^\prime_b~^2P_{\lambda} \frac{1}{2}^- \rangle(6233)~~}$}
&\multicolumn{2}{c}{$\underline{~~|\Xi^\prime_b~^2P_{\lambda} \frac{3}{2}^- \rangle(6234)~~}$}
&\multicolumn{2}{c}{$\underline{~~|\Xi^\prime_b~^4P_{\lambda} \frac{1}{2}^- \rangle(6227)~~}$}
&\multicolumn{2}{c}{$\underline{~~|\Xi^\prime_b~^4P_{\lambda} \frac{3}{2}^- \rangle(6224)~~}$}
&\multicolumn{2}{c}{$\underline{~~|\Xi^\prime_b~^4P_{\lambda} \frac{5}{2}^- \rangle(6226)~~}$} \\
                                 &~~~~~~ $\Gamma_i$         &$\mathcal{B}_i(\%)$ & $~~~~~~\Gamma_i$         & $\mathcal{B}_i(\%)$& $~~~~~~\Gamma_i$  & $\mathcal{B}_i(\%)$
                 &~~~~~~ $\Gamma_i$         & $\mathcal{B}_i(\%)$& $~~~~~~\Gamma_i$         & $\mathcal{B}_i(\%)$\\ \hline
$ \to        \Xi_b^{0}\gamma$         &~~~72.2   &0.27       &~~~72.8  &0.30       &~~~34.0   &0.11      &~~~94.0   &0.59        &~~~47.7  &0.20   \\
$ \to        \Xi_b^{-}\gamma$         &~~~0.0    &0.0        &~~~0.0   &0.0        &~~~0.00   &0.0       &~~~0.0    &0.0         &~~~0.0   &0.0   \\
$ \to  \Xi_b^{\prime0}\gamma$         &~~~76.3   &0.28       &~~~43.9  &0.18       &~~~0.25   &$<0.01$   &~~~0.67   &$<0.01$     &~~~0.44  &$<0.01$  \\
$ \to  \Xi_b^{\prime-}\gamma$         &~~~190    &0.70       &~~~92.3  &0.38       &~~~1.48   &$<0.01$   &~~~2.94   &0.02        &~~~1.88  &$<0.01$   \\
$ \to  \Xi_b^{*\prime0}\gamma$        &~~~0.89   &$<0.01$    &~~~0.90  &$<0.01$    &~~~69.5   &0.22      &~~~47.5   &0.30        &~~~41.5  &0.17  \\
$ \to  \Xi_b^{*\prime-}\gamma$        &~~~3.54   &0.01       &~~~3.60  &0.01       &~~~164    &0.51      &~~~104    &0.66        &~~~88.2  &0.36  \\
\hline\hline
\end{tabular}
\end{center}
\end{table*}

Again, we extend the study of the $1P$-wave $\Xi'_c$ and $\Xi'_b$ states to their radiative decays into the $1S$-wave states. The predicted partial widths (keV) and branching are listed in Table~\ref{raXipp}.
It shows that in the $\Xi^{\prime}_c$ family, the neutral spin-1/2 states $|\Xi^\prime_c~^2P_{\lambda} \frac{1}{2}^- \rangle$ and $|\Xi^\prime_c~^2P_{\lambda} \frac{3}{2}^- \rangle$ have
large decay rates into $\Xi_c^{\prime0}\gamma$ via the $E1$ transition. Their branching fractions are estimated
to be $\mathcal{O}(10^{-2})$. In contrast, the neutral spin-3/2 states $|\Xi^\prime_c~^4P_{\lambda} \frac{1}{2}^- \rangle$, $|\Xi^\prime_c~^4P_{\lambda} \frac{3}{2}^- \rangle$ and $|\Xi^\prime_c~^4P_{\lambda} \frac{5}{2}^- \rangle$ have
large decay rates into $\Xi_c^{\prime*0}\gamma$ via the $E1$ transition, and their branching fractions are estimated
to be $\mathcal{O}(10^{-3})-\mathcal{O}(10^{-2})$.

The radiative decays of the $\Xi'_b$ family are also investigated. It shows that the neutral spin-1/2 states $|\Xi^\prime_b~^2P_{\lambda} \frac{1}{2}^- \rangle$ and $|\Xi^\prime_b~^2P_{\lambda} \frac{3}{2}^- \rangle$ have large radiative decay rates into the $\Xi_b^0 \gamma$ and $\Xi_b^{\prime0} \gamma$ channels, while the charged spin-1/2 states $|\Xi^\prime_b~^2P_{\lambda} \frac{1}{2}^- \rangle$ and $|\Xi^\prime_b~^2P_{\lambda} \frac{3}{2}^- \rangle$ have large radiative decay rates into the $\Xi_b^- \gamma$ channel.
The neutral states spin-3/2 $|\Xi^\prime_b~^4P_{\lambda} \frac{1}{2}^- \rangle$, $|\Xi^\prime_b~^4P_{\lambda} \frac{3}{2}^- \rangle$, and $|\Xi^\prime_b~^4P_{\lambda} \frac{5}{2}^- \rangle$ have large decay rates into the $\Xi_b^0 \gamma$ and $\Xi_b^{\prime*0} \gamma$
channels. The charged spin-3/2 states $|\Xi^\prime_b~^4P_{\lambda} \frac{1}{2}^- \rangle$, $|\Xi^\prime_b~^4P_{\lambda} \frac{3}{2}^- \rangle$ and $|\Xi^\prime_b~^4P_{\lambda} \frac{5}{2}^- \rangle$ have relatively large radiative decay rates into the $\Xi_b^{\prime*-} \gamma$ channel.
The branching fractions for these radiative processes are estimated to be $\mathcal{O}(10^{-3})$.
To establish the missing $1P$-wave states of $\Xi'_c$ and $\Xi'_b$, the experimental investigation of their radiative transitions
seems to be necessary.

\subsection{The $\Omega_c$ and $\Omega_b$ states}

\subsubsection{$1S$ states}

In the $\Omega_c$ family, the ground $1S$-wave states
with $J^P=1/2^+$ and $J^P=3/2^+$, i.e.,
$\Omega_c^0$ and $\Omega_c(2770)^0$ (denoted by $\Omega_c^{*}$) have been established.
The $\Omega_c(2770)^0$ was first observed by
BaBar in the radiative decay channel $\Omega_c^0\gamma$~\cite{Aubert:2006je}.
In our previous work, the partial width of the radiative transition $\Omega_c(2770)\rightarrow\Omega_c\gamma$
is predicted to be $\Gamma[\Omega_c(2770)\rightarrow\Omega_c\gamma]\simeq 0.89$ keV
~\cite{Wang:2017hej}, which is consistent with other model predictions (see Tab.~\ref{Rad}).

For the $\Omega_b$ sector, the $J^P=3/2^+$ state $\Omega_b^{*}$ is still missing.
The radiative decay process $\Omega_b^{*}\to \Omega_b\gamma$
should play an important role in its decays.
Taking the mass $M=6090$ MeV for $\Omega_b^{*}$, we study
this radiative decay process and obtain the partial width
\begin{eqnarray}
\Gamma[\Omega_b^{*}\to\Omega_b\gamma]\simeq 100 \ \mathrm{eV} \ .
\end{eqnarray}
This value is about an order of magnitude smaller than that of $\Omega_c(2770)^0$.
A small partial width of $\Gamma[\Omega_b^{*}\to\Omega_b\gamma]$ was also
predicted in Refs.~\cite{Aliev:2014bma,Wang:2009cd}.
The tiny radiative decay partial width of the radiative transition $\Omega_b^{*}\to\Omega_b\gamma$
may explain why $\Omega_b^{*}$ is still not observed in the $\Omega_b\gamma$ channel.

\begin{table*}[htp]
\begin{center}
\caption{\label{owmigeP}  Partial widths for the strong and radiative decays of the $\Omega_b$ baryons.}
\begin{tabular}{ccccccccccccccccccccccccccc}\hline\hline
       State       &Mass (MeV)~\cite{Ebert:2011kk}   &~~$\Gamma[\Xi_bK]$ (MeV)~~ &~~$\Gamma[\Omega_b\gamma]$ (keV) ~~
       &~~$\Gamma[\Omega_b^*\gamma]$(keV)~~  &~~$\Gamma_{\mathrm{total}}$ (MeV)        \\ \hline
%$|\Omega_b~^2S \frac{1}{2}^+\rangle$             &6064           &                           & & & \\
$|\Omega_b~^4S \frac{3}{2}^+\rangle$             &6088           & $\cdot\cdot\cdot$        &$0.09$  &$\cdot\cdot\cdot$ & $\cdot\cdot\cdot$   \\
$|\Omega_b~^2P_{\lambda} \frac{1}{2}^-\rangle$   &6339           &49.38                &$154        $&$1.49$   &49.53          \\
$|\Omega_b~^2P_{\lambda} \frac{3}{2}^-\rangle$   &6340           &1.82                 &$83.4$       &$1.51$    &1.90       \\
$|\Omega_b~^4P_{\lambda} \frac{1}{2}^-\rangle$   &6330           &94.98                &$0.64$        &$99.23$   &95.08         \\
$|\Omega_b~^4P_{\lambda} \frac{3}{2}^-\rangle$   &6331           &0.22                 &$1.81$        &$70.68$   &0.29         \\
$|\Omega_b~^4P_{\lambda} \frac{5}{2}^-\rangle$   &6334           &1.60                 &$1.21$        &$63.26$   &1.66         \\
%                             $2^2S_{\lambda\lambda} \frac{1}{2}^+$  &6450     &0.51      &0.33     &/        &&     &0.84         \\
%                             $2^4S_{\lambda\lambda} \frac{3}{2}^+$  &6461     &0.36      &0.14     &0.009    &&     &0.509         \\
\hline
\end{tabular}
\end{center}
\end{table*}

\begin{figure}[h]
\begin{center}
\centering  \epsfxsize=7.6 cm \epsfbox{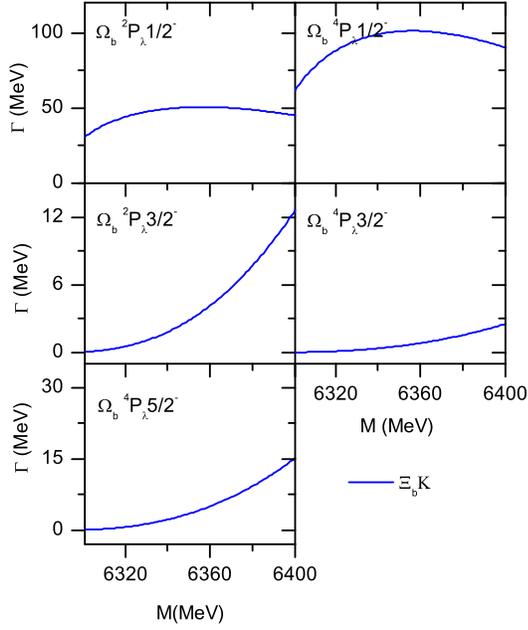} %\epsfxsize=5.0 cm \epsfbox{1DXB.eps}
\vspace{-0.5 cm} \caption{Strong decay widths of the $1P$ states in the $\Omega_b$ family as functions of their mass.  } \label{Omegab}
\end{center}
\end{figure}

\subsubsection{$1P$ states}

In the $\Omega_c$ and $\Omega_b$ families, there are five $1P$-wave $\lambda$-mode excited states,
$|^2P_\lambda\frac{1}{2}^-\rangle$, $|^2P_\lambda\frac{3}{2}^-\rangle$, $|^4P_\lambda\frac{1}{2}^-\rangle$,
$|^4P_\lambda\frac{3}{2}^-\rangle$ and $|^4P_\lambda\frac{5}{2}^-\rangle$.
The typical masses of these $1P$-wave $\Omega_c$ and $\Omega_b$ states
are $\sim 3.05$ and $\sim 6.34$ GeV within various quark model predictions (see Table~\ref{sp1}).

Recently, five new narrow $\Omega_c(X)$ states, $\Omega_c(3000)$, $\Omega_c(3050)$,
$\Omega_c(3066)$, $\Omega_c(3090)$, and $\Omega_c(3119)$, were observed
in the $\Xi_c^{+}K^-$ channel by the LHCb
Collaboration~\cite{Aaij:2017nav}. In Ref.~\cite{Wang:2017hej}, we have studied the strong
and radiative decay properties of these $1P$-wave $\Omega_c$ states.
Combining our predictions with the observations, we suggested that
$\Omega_c$(3050), $\Omega_c$(3066) and $\Omega_c$(3090) may be assigned as the $P$-wave states
$|\Omega_c~^4P_\lambda\frac{3}{2}^-\rangle$, $|\Omega_c~^2P_\lambda\frac{3}{2}^-\rangle$
and $|\Omega_c~^4P_\lambda\frac{5}{2}^-\rangle$, respectively;
while $\Omega_c$(3000) may be explained as the $1P$ mixed state $|\Omega_c~P_\lambda\frac{1}{2}^-\rangle_1$
via the $|^2P_\lambda\frac{1}{2}^-\rangle$-$|^4P_\lambda\frac{1}{2}^-\rangle$ mixing
with a mixing angle $\phi=24^\circ$ or $43^\circ$. Furthermore, the other mixed
state $|^2P_{\lambda}\frac{1}{2}^-\rangle_2$ as the partner of $\Omega_c(3000)$
with a broad width of $\sim 100$ MeV was predicted.

In the $\Omega_b$ family, there are no signals of the $1P$-wave states from experiment.
To provide useful references for observing these missing states in forthcoming experiments,
we study their strong decays within ChQM. Our results are plotted in Fig.~\ref{Omegab}.
It shows that $\Xi_b K$ may be the only OZI-allowed
two-body strong decay channel for these $1P$-wave $\Omega_b$ states.
Taking the masses of the $1P$-wave states predicted
within RQM~\cite{Ebert:2011kk}, we find that the $J^P=3/2^-$ state
$|\Omega_b~^2P_\lambda\frac{3}{2}^-\rangle$ and the $J^P=5/2^-$ state $|\Omega_b~^4P_\lambda\frac{5}{2}^-\rangle$
have compatible decay widths of $\Gamma\simeq 2$ MeV. Again, to distinguish these two states,
more experimental information, such as the angular distributions or radiative decay properties, should be useful.
The decay width of the spin-3/2 state $|\Omega_b~^4P_\lambda\frac{3}{2}^-\rangle$ is about $0.3$ MeV, which is about an
order of magnitude smaller than that of the spin-1/2 state $|\Omega_b~^2P_\lambda\frac{3}{2}^-\rangle$.
The $J^P=1/2^-$ states $|\Omega_b~^2P_\lambda\frac{1}{2}^-\rangle$ and $|\Omega_b~^4P_\lambda\frac{1}{2}^-\rangle$
should be broad states, their widths are predicted to be
$\sim 50$ MeV and $\sim 100$ MeV, respectively. It should be mentioned that very recently,
the strong decays of the first orbitally and radially excited $\Omega_b$ baryons
were studied within the LCQSR~\cite{Agaev:2017nn}. The widths for
$|\Omega_b~^4P_\lambda\frac{3}{2}^-\rangle$ and $|\Omega_b~^2P_\lambda\frac{1}{2}^-\rangle$
were predicted to be $\sim 0.04$ MeV and $\sim 3.97$ MeV, respectively~\cite{Agaev:2017nn},
which are about an order of magnitude smaller than our predictions.

The radiative decays of these
$1P$-wave $\Omega_b$ states are also estimated within the quark model.
Our results are given in Table~\ref{owmigeP}.
For the narrow widths of $|\Omega_b~^2P_\lambda\frac{3}{2}^-\rangle$,
$|\Omega_b~^4P_\lambda\frac{3}{2}^-\rangle$ and $|\Omega_b~^4P_\lambda\frac{5}{2}^-\rangle$,
the branching fractions for the radiative decay processes
$|\Omega_b~^2P_\lambda\frac{3}{2}^-\rangle\to\Omega_b\gamma$,
$|\Omega_b~^4P_\lambda\frac{3}{2}^-\rangle\to \Omega_b^*\gamma$ and
$|\Omega_b~^4P_\lambda\frac{5}{2}^-\rangle\to \Omega_b^*\gamma$ might
reach up to $\mathcal{O}(10\%)$, which seem to be accessible in future experiments. Finally, it should be pointed out that the radiative decays of the spin-1/2 excitation states
$|\Omega_b~^2P_\lambda\frac{1}{2}^-\rangle$ and $|\Omega_b~^2P_\lambda\frac{3}{2}^-\rangle$
are very different from that of the spin-3/2 excitations $|\Omega_b~^4P_\lambda\frac{1}{2}^-\rangle$, $|\Omega_b~^4P_\lambda\frac{3}{2}^-\rangle$ and
$|\Omega_b~^4P_\lambda\frac{5}{2}^-\rangle$. This feature will be helpful for distinguishing the spin-1/2 and -3/2 excitation states.

\section{Summary}\label{suma}

A systematic study of the strong and radiative decays of the low-lying
$S$- and $P$-wave singly heavy baryons in a constituent quark model is presented in this work. Although there still lack experimental data for a better understanding of the heavy baryon spectra, we find that useful information about the heavy baryon structures can still be extracted by combining their strong and radiative decays. Several key results from this study can be learned here,

\begin{itemize}

\item For the neutral partners of the ground state heavy baryons, $\Sigma_b^{0}$ and $\Sigma_b^{*0}$, their radiative transitions to $\Lambda_b^0\gamma$ can have a branching fraction of $\mathcal{O}(10^{-2})$ and is an ideal channel for establishing them in experiments.

\item For the $1P$-wave $\Xi_b$ states of
$\bar{\mathbf{3}}_F$, i.e., $|\Xi_b~^2P_\lambda \frac{1}{2}^-\rangle$
and $|\Xi_b~^2P_\lambda \frac{3}{2}^-\rangle$, they appear to have very narrow
widths and are dominated by the $\Xi_b'\pi$ and $\Xi_b^{'*}\pi$ decay channels, respectively. Also, they have sizeable radiative
decay rates into $\Xi_b^{0,-} \gamma$.

\item For the $1P$-wave states of $\mathbf{6}_F$, there exist rather different decay properties within the multiplets, i.e. $|^2P_{\lambda}\frac{1}{2}^- \rangle$, $|^4P_{\lambda} \frac{1}{2}^- \rangle$, $|^2P_{\lambda} \frac{3}{2}^- \rangle$, $|^4P_{\lambda} \frac{3}{2}^- \rangle$, and $|^4P_{\lambda} \frac{5}{2}^- \rangle$. We show that some of these states in the  $\Sigma_{c(b)}$ and $\Xi'_{c(b)}$ families are quite narrow and dominated by their single meson hadronic decays. In particular, the $|\Omega_b~^4P_{\lambda} \frac{3}{2}^- \rangle$ might be very narrow and has a width of a few hundred keV decaying into the $\Xi_bK$ channel.

\item We identify some of those radiative decay channels for the $1P$-wave states which can be helpful for future searches of their signals in experiments. For instance, the radiative transitions $|\Sigma_c^{++(0)}~^2P_{\lambda} \frac{3}{2}^- \rangle\to\Sigma_c^{++(0)}\gamma$
and $|\Sigma_c^{++(0)}~^4P_{\lambda} \frac{5}{2}^- \rangle\to\Sigma_c^{*++(0)}\gamma$ have large decay rates up to $\mathcal{O}(10^{-2})$. This happens to other multiplets of $|^2P_{\lambda} \frac{3}{2}^- \rangle$ and $|^4P_{\lambda} \frac{5}{2}^- \rangle$ in the $\Sigma_b^+$, $\Xi_c^{\prime 0}$, $\Xi_b'$ and $\Omega_b$ spectra.

\end{itemize}

In brief, we find that the ChQM can still serve as a useful tool for investigating the heavy baryon mesonic decays and radiative transitions. Some of the featured results are consistent with other model approaches. These results could be helpful for future experimental search for some of those excited states and provide deeper insights into our understanding of the heavy baryon spectroscopy.

\section*{  Acknowledgments }

This work is supported, in part, by the National Natural Science Foundation of China under Grants No. 11375061, No. 11775078,
No. 11425525, No. 11521505, and No. 11261130311;
and National Key Basic Research Program of China under Contract No. 2015CB856700.
%\appendix
%%%%%%%%%%%%%%%%%%%%%%%%%%%%%%%%%%%%%%%%%%%%%%%%%%%%%%%%%%%%%%%%%%555

\bibliographystyle{unsrt}

\end{document}